\documentclass[10pt]{article}
\usepackage{amsmath,amsthm,latexsym,amssymb,amsfonts,epsfig,psfrag}
\usepackage{cite}
\usepackage{color}
\usepackage{authblk}
\makeatletter \@addtoreset{equation}{section} \makeatother

\def\be{\begin{equation}}
\def\ee{\end{equation}}
\def\ba{\begin{eqnarray}}
\def\ea{\end{eqnarray}}

\newcommand\nn{\nonumber}
\newcommand\q{\quad}
\def\Nl{{\mathchoice
{\setbox0=\hbox{$\displaystyle\rm N$}\hbox{\hbox to0pt
{\kern0.4\wd0\vrule height0.9\ht0\hss}\box0}}
{\setbox0=\hbox{$\textstyle\rm N$}\hbox{\hbox to0pt
{\kern0.4\wd0\vrule height0.9\ht0\hss}\box0}}
{\setbox0=\hbox{$\scriptstyle\rm N$}\hbox{\hbox to0pt
{\kern0.4\wd0\vrule height0.9\ht0\hss}\box0}}
{\setbox0=\hbox{$\scriptscriptstyle\rm N$}\hbox{\hbox to0pt
{\kern0.4\wd0\vrule height0.9\ht0\hss}\box0}}}}
\def\Zl{{\mathchoice
{\setbox0=\hbox{$\displaystyle\rm Z$}\hbox{\hbox to0pt
{\kern0.4\wd0\vrule height0.9\ht0\hss}\box0}}
{\setbox0=\hbox{$\textstyle\rm Z$}\hbox{\hbox to0pt
{\kern0.4\wd0\vrule height0.9\ht0\hss}\box0}}
{\setbox0=\hbox{$\scriptstyle\rm Z$}\hbox{\hbox to0pt
{\kern0.4\wd0\vrule height0.9\ht0\hss}\box0}}
{\setbox0=\hbox{$\scriptscriptstyle\rm Z$}\hbox{\hbox to0pt
{\kern0.4\wd0\vrule height0.9\ht0\hss}\box0}}}}
\def\Ql{{\mathchoice
{\setbox0=\hbox{$\displaystyle\rm Q$}\hbox{\hbox to0pt
{\kern0.4\wd0\vrule height0.9\ht0\hss}\box0}}
{\setbox0=\hbox{$\textstyle\rm Q$}\hbox{\hbox to0pt
{\kern0.4\wd0\vrule height0.9\ht0\hss}\box0}}
{\setbox0=\hbox{$\scriptstyle\rm Q$}\hbox{\hbox to0pt
{\kern0.4\wd0\vrule height0.9\ht0\hss}\box0}}
{\setbox0=\hbox{$\scriptscriptstyle\rm Q$}\hbox{\hbox to0pt
{\kern0.4\wd0\vrule height0.9\ht0\hss}\box0}}}}
\def\Rl{{\mathchoice
{\setbox0=\hbox{$\displaystyle\rm R$}\hbox{\hbox to0pt
{\kern0.4\wd0\vrule height0.9\ht0\hss}\box0}}
{\setbox0=\hbox{$\textstyle\rm R$}\hbox{\hbox to0pt
{\kern0.4\wd0\vrule height0.9\ht0\hss}\box0}}
{\setbox0=\hbox{$\scriptstyle\rm R$}\hbox{\hbox to0pt
{\kern0.4\wd0\vrule height0.9\ht0\hss}\box0}}
{\setbox0=\hbox{$\scriptscriptstyle\rm R$}\hbox{\hbox to0pt
{\kern0.4\wd0\vrule height0.9\ht0\hss}\box0}}}}
\def\Cl{{\mathchoice
{\setbox0=\hbox{$\displaystyle\rm C$}\hbox{\hbox to0pt
{\kern0.4\wd0\vrule height0.9\ht0\hss}\box0}}
{\setbox0=\hbox{$\textstyle\rm C$}\hbox{\hbox to0pt
{\kern0.4\wd0\vrule height0.9\ht0\hss}\box0}}
{\setbox0=\hbox{$\scriptstyle\rm C$}\hbox{\hbox to0pt
{\kern0.4\wd0\vrule height0.9\ht0\hss}\box0}}
{\setbox0=\hbox{$\scriptscriptstyle\rm C$}\hbox{\hbox to0pt
{\kern0.4\wd0\vrule height0.9\ht0\hss}\box0}}}}
\def\Hl{{\mathchoice
{\setbox0=\hbox{$\displaystyle\rm H$}\hbox{\hbox to0pt
{\kern0.4\wd0\vrule height0.9\ht0\hss}\box0}}
{\setbox0=\hbox{$\textstyle\rm H$}\hbox{\hbox to0pt
{\kern0.4\wd0\vrule height0.9\ht0\hss}\box0}}
{\setbox0=\hbox{$\scriptstyle\rm H$}\hbox{\hbox to0pt
{\kern0.4\wd0\vrule height0.9\ht0\hss}\box0}}
{\setbox0=\hbox{$\scriptscriptstyle\rm H$}\hbox{\hbox to0pt
{\kern0.4\wd0\vrule height0.9\ht0\hss}\box0}}}}
\def\Ol{{\mathchoice
{\setbox0=\hbox{$\displaystyle\rm O$}\hbox{\hbox to0pt
{\kern0.4\wd0\vrule height0.9\ht0\hss}\box0}}
{\setbox0=\hbox{$\textstyle\rm O$}\hbox{\hbox to0pt
{\kern0.4\wd0\vrule height0.9\ht0\hss}\box0}}
{\setbox0=\hbox{$\scriptstyle\rm O$}\hbox{\hbox to0pt
{\kern0.4\wd0\vrule height0.9\ht0\hss}\box0}}
{\setbox0=\hbox{$\scriptscriptstyle\rm O$}\hbox{\hbox to0pt
{\kern0.4\wd0\vrule height0.9\ht0\hss}\box0}}}}
\newcommand{\cc}{\mathcal C}
\newcommand{\cd}{\mathcal D}

\newcommand{\cf}{\mathcal F}

\newcommand{\ch}{\mathcal H}

\newcommand{\cp}{\mathcal P}

\newcommand{\cs}{\mathcal S}
\newcommand{\ct}{\mathcal T}

\newcommand{\p}{\partial}
\newcommand{\f}{\frac}

\begin{document}


\title{Chaos, Dirac observables and constraint quantization}

\author[1]{Bianca Dittrich\thanks{\texttt{bdittrich@perimeterinstitute.ca}}}
\author[1]{Philipp A.\ H\"ohn\thanks{\texttt{phoehn@perimeterinstitute.ca}}}
\author[2,3]{Tim A.\ Koslowski\thanks{\texttt{t.a.koslowski@gmail.com}}}
\author[1,4,5]{Mike I.\ Nelson\thanks{\texttt{mike@aims.edu.gh}}}
\affil[1]{Perimeter Institute for Theoretical Physics, 31 Caroline Street North, Waterloo, ON N2L 2Y5, Canada}
\affil[2]{University of New Brunswick, Fredericton, NB, E3B 5A3 Canada}
\affil[3]{Instituto de Ciencias Nucleares, Universidad Nacional Aut\'onoma de M\'exico,
Apartado Postal 70-543, M\'exico D.F. 04510, M\'exico}
\affil[4]{African Institute for Mathematical Sciences
P.O Box LG 197, Legon, Accra, Ghana}
\affil[5]{Department of Mathematics,
University of Ghana, P.O. Box LG 62, Legon, Accra, Ghana}

\date{}


\setcounter{footnote}{0}
\maketitle

\vspace{-.9cm}

\begin{abstract}
There is good evidence that full general relativity is non-integrable or even chaotic. We point out the severe repercussions: differentiable Dirac observables and a reduced phase space do not exist in non-integrable constrained systems and are thus unlikely to occur in a generic general relativistic context. Instead, gauge invariant quantities generally become discontinuous, thus not admitting Poisson-algebraic structures and posing serious challenges to a quantization. Non-integrability also renders the paradigm of relational dynamics cumbersome, thereby straining common interpretations of the dynamics. We illustrate these conceptual and technical challenges with simple toy models. In particular, we exhibit reparametrization invariant models which fail to be integrable and, as a consequence, can either not be quantized with standard methods or lead to sick quantum theories without a semiclassical limit. These troubles are qualitatively distinct from semiclassical subtleties in unconstrained quantum chaos and can be directly traced back to the scarcity of Dirac observables. As a possible resolution, we propose to change the method of quantization by refining the configuration space topology until the generalized observables become continuous in the new topology and can acquire a quantum representation. This leads to the polymer quantization method underlying loop quantum cosmology and gravity. Remarkably, the polymer quantum theory circumvents the problems of the quantization with smooth topology, indicating that non-integrability and chaos, while a challenge, may not be a fundamental obstruction for quantum gravity. 
\end{abstract}


\section{Introduction}

In order to extract physical information from gauge theories, one either constructs gauge invariant observables or fixes a gauge. In a canonical language, this usually means either constructing functions which Poisson-commute on the constraint surface with the gauge generating first class constraints or to fix the flow of the latter by singling out points within each gauge orbit \cite{Dirac,Henneaux:1992ig}. Such gauge invariant observables are thus the `constants of motion' of the gauge generators and referred to as Dirac observables. In the quantum theory, one typically deals with gauge symmetries by canonically quantizing according to the Dirac (and sometimes even the reduction) method or by performing a Faddeev-Popov type gauge fixing in the corresponding path integral formulation. In either method, Dirac observables assume a pivotal role in the interpretation of the quantum theory.  The reason for this is that the operators on the physical Hilbert space, obtained after solving the (quantum) constraints, are the quantum Dirac observables.
Also, if one uses a path integral quantization, the path integral acts as a projector onto the physical Hilbert space \cite{Halliwell:1990qr,Rovelli:1998dx}, and (quantum) Dirac observables are again essential to characterize the resulting quantization.

Among the gauge theories, general relativity possesses special attributes: it features spacetime diffeomorphism symmetry and, in its canonical formulation, is a totally constrained system \cite{Dirac:1958sc,Arnowitt:1962hi,Thiemann:2007zz}, which means that also its dynamics is generated by a (first class) Hamiltonian constraint. This has deep consequences: Firstly, diffeomorphism invariant observables are necessarily non-local in a spacetime manifold sense (i.e., they must involve spatial derivatives of infinitely high order) \cite{Torre:1993fq} which poses challenges to connecting them to the (background dependent) local observables of quantum field theory \cite{Giddings:2005id,Giddings:2015lla}. Secondly, gravitational Dirac observables are coordinate and, in particular, time coordinate independent; they are `constants of motion' of the Hamiltonian constraint, thus appearing to be non--dynamical, which leads to the infamous `problem of time' in the quantum theory \cite{Kuchar:1991qf,Isham:1992ms,Anderson:2012vk}. 

A possible resolution to the seemingly non-local and non-dynamical character of gravitational observables is often hoped to be found within the relational nature of general relativity itself: localization and evolution of systems and dynamical quantities are only meaningful relative to other dynamical quantities -- rather than with respect to a background. Gravitational observables capturing the relative localizations and dynamics should thus be relational objects that correlate different dynamical quantities \cite{relrov,Rovelli:1989jn,Rovelli:1990jm,Rovelli:1990ph,Rovelli:1990pi,Rovelli:2004tv,Tambornino:2011vg}. Such relational Dirac observables are often also called `evolving constants of motion'. Classically, much progress has been made in defining and, at least formally, constructing relational observables \cite{Dittrich:2004cb,Dittrich:2005kc,Tambornino:2011vg,Hajicek:1994py,Hajicek:1995en,Hajicek:1996xk,Anderson:2013wua}. In particular, there now exist perturbative expansions, applicable to fixed backgrounds and to perturbations around symmetry reduced sectors (e.g.\ in cosmology), for constructing approximate observables, which in their lowest order approximation reproduce the local observables of field theory on a fixed background \cite{Dittrich:2006ee,Dittrich:2007jx}. On the other hand, much work remains to be done in the quantum theory -- for both the finite dimensional and the field theory cases -- where no general method for representing the quantum analogues of relational observables exists. As a possible remedy, effective techniques for evaluating and constructing relational observables in the semiclassical limit of finite dimensional systems have been developed which, in principle, can also be generalized beyond the semiclassical regime and to the field theory case \cite{Bojowald:2010xp,Bojowald:2010qw,Hohn:2011us,Hoehn:2011jw}.

Unfortunately, gravitational Dirac observables (relational or otherwise) are notoriously hard to find explicitly as their construction requires solving the dynamics (see \cite{Tambornino:2011vg,Anderson:2013wua} for recent reviews). In fact, for full general relativity no Dirac observables are known with the exception of the ADM charges in asymptotically flat spacetimes \cite{Arnowitt:1962hi}, the boundary charges in asymptotically anti de Sitter spacetimes \cite{Henneaux:1984xu,Henneaux:1985tv,Anderson:1996sc} or, perhaps, the deparametrized relational observables in dust filled spacetimes \cite{Brown:1994py,Husain:2011tk}. However, these exceptions do not account for a generic situation in full general relativity since the first two examples require special asymptotic symmetries, while the third relies on a specially engineered matter content. Indeed, Anderson and Torre showed that, except for the asymptotic symmetries, there do not exist any further (generalized) symmetries in full general relativity \cite{Torre:1993jm}. 


Relational Dirac observables, in particular, are difficult to construct and evaluate explicitly because they require internal time or `clock' variables which parametrize the orbits of the gauge generating constraints. Since fixing the `clocks' to a particular set of values should single out a point on each gauge orbit, the existence of such `clock' variables is directly connected to the existence of good gauge fixing conditions for the flows of the constraints. Indeed, relational observables can be viewed as gauge invariant extensions of gauge restricted quantities \cite{Henneaux:1992ig,Dittrich:2004cb,Dittrich:2005kc}. In complete analogy to the Gribov problem in non-abelian gauge theories, however, there will generically exist global obstructions to the existence of ideal `clocks' whose level surfaces are intersected precisely once by every classical trajectory; this has become known as the `global problem of time' \cite{Kuchar:1991qf,Isham:1992ms,Hajicek:1994py,Hajicek:1995en,Hajicek:1988he,Schon:1989pe,Hajicek:1989ex,Rovelli:1990jm,Bojowald:2010xp,Bojowald:2010qw,Hohn:2011us,Hoehn:2011jw}. Furthermore, a generic `clock' variable will couple to the `evolving' degrees of freedom which severely complicates the resolution of the quantum relational dynamics \cite{Giddings:2005id,Hohn:2011us,Marolf:1994nz,Marolf:2009wp}. In other words, ideal `clocks' will in general not exist.
This is emphasized in the work by Torre \cite{Torre-notpara}, showing that general relativity is not a parametrized system, i.e.\ there does not exist an ideal (set of) clock variables. This is an important point since -- as we shall see -- the problem we are discussing here does not arise for parametrized systems.

In (loop) quantum cosmology, one usually turns a blind eye to this challenge by employing highly idealized and decoupled matter `clocks' such as a free, massless scalar \cite{Bojowald:2008zzb,Ashtekar:2011ni} or dust field \cite{Husain:2011tm}, or by resorting to model specific geometric internal time variables \cite{MartinBenito:2009qu}. While this is useful for gaining insights into non-generic quantum cosmological models, it clearly does not capture the qualitative features of relational dynamics and observables in a generic cosmological model. The effective techniques \cite{Bojowald:2010xp,Bojowald:2010qw,Hohn:2011us,Hoehn:2011jw} for evaluating quantum relational dynamics have been developed precisely to remedy the restriction to special `clocks' and, instead, to generalize to arbitrary degrees of freedom as internal times. In particular, this method permits one (a) to switch the `clock' in the quantum theory, thereby enabling one to evolve past `clock' pathologies by patching together evolutions in different internal times, and (b) to construct and give meaning to transient/local observables, both of which seem indispensable in the light of the `global problem of time'. 

However, as argued in \cite{Hohn:2011us},\footnote{See especially Sec.\ IV C 4 of \cite{Hohn:2011us}.} the `global problem of time' will become unsurmountable in the presence of chaos. Chaos generically prevents the existence of differentiable internal time variables which parametrize the orbits of the first class constraints (even locally) in every open neighbourhood of the constraint surface. (It does not prevent the existence of discontinuous internal time variables which, however, cannot be inserted in Poisson brackets or the existence of local `clocks' which parametrize the orbits away from chaotic regions.) This has severe repercussions for quantum relational dynamics which, even semiclassically, seems to break down generically 
\cite{Hohn:2011us}. The absence of a good relational clock also implies that the Gribov problem becomes generically untractable (at least in some neighbourhoods of the constraint surface) in the presence of chaos; a given gauge might either miss `most' trajectories and/or have densely many solutions (see also the main body of this article). Since chaos features in generic dynamical systems \cite{ottchaos,gutzwillerbook,berrychaos}, the understanding of relational dynamics and observables, gained from simple, solvable cosmological models, will thus, in fact, be misleading for a general cosmological model, let alone full quantum gravity. Under such circumstances, it seems advantageous to consider physical transition amplitudes (involving the projector onto the constraints) between kinematical states \cite{Rovelli:2004tv,Rovelli:2001bz}, instead of the more cumbersome relational dynamics. Indeed, such transition amplitudes are more general and, in principle, admit a useful dynamical interpretation also when relational dynamics breaks down. 

This article is devoted to better understanding the repercussions of chaos -- or, more generally, non-integrability -- for the existence of Dirac observables, for relational dynamics and the quantization of constrained systems. We shall argue that, as a consequence of non-integrability which implies the absence of sufficiently many smooth `constants of motion', it is not merely a technical problem which inhibits the construction of Dirac observables for full general relativity or even a generic cosmological model, but a problem in principle: {\it Dirac observables for full general relativity (without special asymptotic symmetries or matter content) almost certainly do not exist}. Before expanding on this in the main body and until then causing an unnecessary confusion, we emphasize that this does not imply that gauge (i.e., diffeomorphism) invariant quantities generically do not exist. Quite the contrary: gauge invariant functions on the constraint surface which are constant on the gauge orbits will {\it always} exist.\footnote{{Such functions can, for instance, be defined by averaging non--invariant functions over gauge orbits.,}} However, these will generically be {\it discontinuous} (or singular, see \cite{UnruhWald,Hajicek:1995en,Hajicek:1996xk} and main body of this article), i.e., non-differentiable and thus not Dirac observables in the literal sense since they cannot be inserted into Poisson brackets with other observables. That is, these generalized gauge invariant quantities will not admit any Poisson algebraic structure.  We emphasize that with the term `Dirac observable' we refer to phase space functions which are smooth (or sufficiently differentiable in all directions) on the constraint surface.

What might appear as an innocent observation, has deep consequences, in particular for the quantum theory: Firstly, it implies that a reduced phase space does not exist under such circumstances since the Poisson (or rather Dirac) bracket between any two functions on the reduced phase space -- if it existed -- are equivalent to a Poisson bracket between two Dirac observables on the constraint surface \cite{Henneaux:1992ig}. But such differentiable Dirac observables are likely not to exist generically. This suggests that {\it a reduced phase space for full general relativity (without special asymptotic symmetries or matter content) almost certainly does not exist} either. 

Secondly, in the absence of Dirac observables one lacks an algebraic structure which can be promoted into the quantum theory. This directly rules out reduced quantization. While the failure of the reduction method by itself is not a great surprise, the absence of Dirac observables and algebraic relations also poses a challenge to a Dirac or path integral quantization through standard methods (if possible at all) because it will become very non-trivial to relate the observables on a physical Hilbert space to those of the classical theory. This severely complicates a semiclassical limit and the interpretation of the quantum theory.

Thirdly, as we shall illustrate by means of toy models, the absence of sufficiently many Dirac observables indeed seems to generically prevent the very existence of a semiclassical limit -- both in terms of physical states and transition amplitudes -- of a quantum theory obtained by a standard Dirac procedure. This supports earlier reports in the literature of an absence of good semiclassical states in a chaotic cosmological model \cite{Kiefer:1988tr,Hohn:2011us}. We emphasize at this point already that these troubles are qualitatively and conceptually distinct from the usual semiclassical subtleties in unconstrained quantum chaos where typically at least classically periodic orbits can be reproduced and wave packets have at least a short term coherence \cite{ottchaos,gutzwillerbook,berrychaos}. In the present constrained case, the absence of semiclassical states is directly tied to the absence of sufficiently many Dirac observables. Such Dirac quantized models can generically not reproduce any of the classical dynamics and may thus be considered ill.\footnote{This is also distinct from systems such as a hydrogen atom where states of low energy do not reproduce the (wrong) classical low energy dynamics, yet highly energetic states come arbitrarily close to the classical high energy case.} In particular, we shall also exhibit a system which has meaningful classical dynamics, yet can neither be quantized through the reduced nor the standard Dirac method, thereby providing an extreme example for a sick quantum theory (and, in particular, the absence of a semiclassical limit).

All these observations seem, at first sight, to be rather devastating news for the quantization of non-integrable constrained systems in general and general relativity in particular. On a second look, however, these properties turn out to be less dramatic and do not necessarily constitute a fundamental road block to quantum gravity. Indeed, one has to remember that quantizing a theory is a highly ambiguous procedure. These findings simply force us to change our preferred method -- and, more precisely, topology -- for quantization. The mentioned troubles with the standard techniques can be directly traced back to the discontinuity of the observables which impedes their quantum representations. However, the notion of (dis)continuity requires a topology and the generalized observables are discontinuous in the standard (Euclidean metric) topology. The key idea we shall propose here for resolving these troubles and usefully translating these observables into a quantum theory will be {\it to refine the topology until the generalized observables become continuous in the new topology.} 

This leads to the good news of this article: Introducing such a refined topology leads to an interesting quantum theory with sufficiently many observables in all presented examples where the standard methods produce a failure. 

Clearly, there is a price to pay for curing the illnesses of standard quantization methods: (a) One has to deal with an unfamiliar discrete topology and the ensuing technical consequences, (b) one obtains a novel parameter in the quantum theory measuring the step size of translation operators which requires a physical interpretation, and (c) the spectra of the conjugate observables will become (Bohr) compactified. However, there already exists a well-developed machinery for dealing with this alternative quantization method. Namely, interestingly, this method is precisely the quantization method of loop quantum cosmology \cite{Bojowald:2008zzb,Ashtekar:2011ni,Ashtekar:2003hd}, sometimes known as `polymer quantization' \cite{Hossain:2010wy,Laddha:2010hp,Corichi:2007tf}, and also appears in a recent novel representation of loop quantum gravity \cite{Bahr:2015bra}. In particular, in loop quantum cosmology, the novel parameter (b) introduced through quantization is closely related to the Planck length and thus carries an immediate physical interpretation. The results presented in this manuscript are a remarkable success of polymer quantization and provide evidence that this method might generally resolve the challenges posed by non-integrability, although the task of finding a useful interpretation of the quantum dynamics (relational or in terms of transition amplitudes) remains.

Given the importance of Dirac observables and relational dynamics for constrained systems and their quantization in general -- and general relativity in particular -- and given the deep consequences of non-integrability featuring in generic systems, it is surprising that this issue has been largely ignored in the recent literature. The exception are older works \cite{UnruhWald,Kuchar:1993ne,Torre:1993jm,Hajicek:1995en,Hajicek:1996xk,Anderson:1995tu,Smolin:2000vs} which, however, only briefly touch on the existence of classical Dirac observables\footnote{In the older literature, Dirac observables are sometimes also referred to as `perennials'.} and 
{are rather brief on}
  the consequences for the quantum theory.\footnote{In particular \cite{UnruhWald} notes that (smooth) Dirac observables will not exist for an ergodic dynamics. Our examples will also be ergodic at least in configuration space, resulting in the lack of sufficiently many Dirac observables.} The first concrete analysis of the deep ramifications of chaos for the quantum theory of constrained systems and relational dynamics in particular has only been carried out recently \cite{Hohn:2011us}. However, in our opinion, the consequences of non-integrability for quantum gravity cannot be underestimated and warrant further study. In this spirit, we hope that the present work opens up a new avenue of research on chaos and non-integrability in (quantum) gravity.\footnote{See also \cite{Haggard:2012jp} for an interesting application of chaos and random matrix theory in quantum gravity to provide evidence that the spectrum of the volume operator features a gap at the zero eigenvalue.}

The remainder of this manuscript is organized as follows. In section \ref{sec_nonint}, we shall make the notion of non-integrability more precise and discuss the evidence for non-integrability in general relativity. In order to display a concrete analysis of the consequences of non-integrability, we introduce a toy model in section \ref{sec_ex1} with ergodic dynamics in configuration space. This model, while not being fully non-integrable or chaotic, lacks sufficiently many Dirac observables and is shown not to possess a reduced phase space. The advantage of this simple `semi-integrable' model is that it can still be treated analytically (in contrast to most chaotic models), yet features generic behaviour of non-integrable models. We quantize this model by means of the Dirac method and standard topology in section \ref{sec_standard}, exhibiting its failure to produce a good semiclassical limit. Thereupon, in section \ref{sec_discrete}, we refine the topology and polymer quantize the same model, showing how an interesting quantum theory with sufficiently many observables arises. In section \ref{sec_furtherex}, we discuss three further twists on this toy model, illustrating various degrees of integrability (however, none of which leads to full non-integrability). In particular, in section \ref{sec_screwedup}, we discuss a model whose classical trajectories are all ergodic and which is shown to prevent a quantization by standard methods, yet which permits a polymer quantization. We conclude with a discussion in section \ref{conc}.

\section{Non-integrability and Dirac observables}\label{sec_nonint}

It is useful to review the concepts of integrability and non-integrability in unconstrained Hamiltonian systems. We shall do so for finite dimensional autonomous systems. 

Consider a system described by a $2N$-dimensional phase space $\cp$ and time independent Hamiltonian $H$. The system is called {\it integrable} if there exist $N$ independent\footnote{That is, the differentials of the $F_i$ are linearly independent everywhere on phase space.} constants of motion $\cf=\{F_1,\ldots,F_N\}$ on $\cp$ with $H\in\cf$ which are in involution
\ba
\{F_k,F_l\}=0\label{involution}
\ea
and thus, by Frobenius' theorem, integrate to a $N$-dimensional submanifolds $M_F$ of $\cp$, determined by the values of $F_1,\ldots,F_N$. Under mild conditions, one can show that $M_F\simeq\mathbb{T}^k\times\mathbb{R}^{N-k}$ where $\mathbb{T}^k$ is a $k$-dimensional torus \cite{arnold2007mathematical}. That is, in an integrable system, the classical trajectories lie on $N$-dimensional submanifolds $M_F$ of phase space.

By contrast, an unconstrained Hamiltonian system is called {\it non-integrable} if no smooth (or differentiable) constants of motion exist globally on $\cp$ which are independent of $H$. In this case, trajectories are not confined to an $N$-dimensional submanifold of $\cp$, but only to a $(2N-1)$-dimensional energy surface; a typical trajectory will explore a region of co-dimension $1$ in phase space. Non-integrability features in generic unconstrained Hamiltonian systems \cite{ottchaos,gutzwillerbook,berrychaos} and there exist concrete theorems stating precise conditions for the absence of smooth constants of motion \cite{arnold2007mathematical}. There are various characterizations of non-integrable systems \cite{ottchaos,gutzwillerbook,berrychaos,arnold2007mathematical}; e.g., they may be ergodic or chaotic. Ergodic or chaotic systems are non-integrable, however, the converse is not necessarily true. The precise definition of chaos or ergodicity are unimportant for our purposes. What matters is the more general notion of non-integrability and the associated absence of constants of motion.

In analogy to these notions for unconstrained Hamiltonian systems, we shall now define three distinct notions of (non-)integrability for constrained systems. Again, we restrict to the finite dimensional case; the field theory case might be defined analogously in terms of local degrees of freedom. Consider a system described by a $2N$-dimensional kinematical phase space $\cp$ and $m_1$ irreducible first class and $2\,m_2$ irreducible second class constraints such that -- for a solvable system -- one would expect $2(N-m_1-m_2)$ independent physical degrees of freedom. Denote by $\cc$ the constraint surface and by $\gamma$ a (copy of the) gauge orbit. 

We begin with a physically somewhat superfluous notion of (non-)integrability, however, mention it for technical completeness.
\begin{description}
\item[strong (non-)integrability:] A constrained system shall be referred to as {\it strongly integrable} if there exist $N$ independent functions $F_1,\ldots,F_N$ on $\cp$, containing the constraints, which are in involution (\ref{involution}) in terms of strong equalities. These functions, again, integrate to $N$-dimensional hypersurfaces in the kinematical phase space. 

A constrained system shall be called {\it strongly non-integrable} if there do not exist any functions, independent of the constraints, which strongly commute with the constraints and if the constraints themselves do not commute strongly.
\end{description}
Clearly, a system with second class constraints can never be strongly integrable. Regardless, strong equalities are usually physically not very interesting for constrained systems such that we shall not further elaborate on this case.

Instead, of most interest to us will be the notion of (non-)integrability in terms of the physically relevant weak inequalities. This is also the notion pertinent to general relativistic systems.
\begin{description}
\item[weak (non-)integrability:] A constrained system will be called {\it weakly integrable} if (a) there exist $2(N-m_1-m_2)$ independent Dirac observables which are also independent of the constraints, and (b) $N-m_1-m_2$ of the Dirac observables are weakly in involution. (a) implies that the system is {\it reducible}, i.e.\ (each connected component of) $\cc/\gamma$ constitutes a reduced phase space $\cp_R$ with symplectic structure (see the comments in the introduction and \cite{Henneaux:1992ig}). Moreover, (b) implies that, together with the $m_1$ first class constraints, the $N-m_1-m_2$ Dirac observables in weak involution form a set of $N-m_2$ functions in weak involution. By Frobenius' theorem, these integrate to $(N+m_2)$-dimensional submanifolds of $\cc$. If the generator of dynamics is first class (this is generically the case \cite{Henneaux:1992ig}), a classical trajectory will remain restricted to such a submanifold. 

On the other hand, a constrained system will be called {\it weakly non-integrable} if there exist no Dirac observables other than the constraints. As argued in the introduction, this entails that a weakly non-integrable system is also necessarily irreducible and does not possess a reduced phase space (see also \cite{Hajicek:1995en,Hajicek:1996xk} for a related discussion of irreducibility). A Marsden-Weinstein type symplectic reduction thus fails in such systems. Furthermore, a classical trajectory will generically not remain restricted to a $(N+m_2)$-dimensional submanifold of $\cc$, instead exploring regions of larger dimension.
\end{description}
We emphasize that the weak notion of integrability is indeed mathematically weaker than the strong notion as it requires less stringent conditions.

Finally, there is an obvious notion of (non-)integrability for reducible systems.
\begin{description}
\item[reduced (non-)integrability:] A constrained system shall be termed {\it reducibly integrable} if the system is (1) weakly integrable and thus reducible, and (2) if the reduced system is integrable on the reduced phase space $\cp_R$, i.e.\ if there exist $N_R$ functions $F_1,\ldots,F_{N_R}$ in involution with respect to the Dirac bracket on the $2N_R$-dimensional $\cp_R$. 

The constrained system will be referred to as {\it reducibly non-integrable} if the system is (1) weakly integrable and thus reducible, and (2) if the reduced system is non-integrable on $\cp_R$, i.e.\ there exist no smooth constants of motion of the surviving physical Hamiltonian on $\cp_R$ (which are independent of this Hamiltonian). 
\end{description}

For example, consider a non-integrable or even chaotic unconstrained Hamiltonian system subject to a (time independent) Hamiltonian $H$ on a $2N$-dimensional phase space $\cp$. Promoting the time $t$ to a dynamical variable and thus extending the phase space to a $2(N+1)$-dimensional phase space produces a system governed by the Hamiltonian constraint $C=p_t+H$. This system is clearly reducible because $t=const$ constitutes a globally valid gauge fixing condition which gives rise to a global Dirac bracket. However, since the reduced phase space is simply the original unextended phase space, $\cp_R\equiv\cp$, the system is clearly reducibly non-integrable. (See also \cite{Hajicek:1995en,Hajicek:1996xk} for a related discussion.) On the other hand, if the unconstrained system was integrable, the parametrized analogue of it would evidently be reducibly integrable.

In the sequel, we shall focus almost exclusively on weak (non-)integrability as this implies consequences for the existence of Dirac observables, the existence of a reduced phase space and quantization. By contrast, we emphasize that there is no fundamental problem in quantizing reducible systems with standard techniques either by the Dirac or reduction method as they possess a reduced phase space and thus sufficiently many Dirac observables which generally admit algebraic representations in the quantum theory. Clearly, it might be technically difficult to solve the quantum dynamics of reducibly non-integrable systems but there is no problem in principle. For instance, for parametrized non-integrable systems the physical Hilbert space is isomorphic to the Hilbert space of the original unparametrized system. It will be a challenge to exactly solve the Schr\"odinger (constraint) equation, but there exist useful approximation techniques and other powerful methods to extract physical information from such systems \cite{ottchaos,gutzwillerbook,berrychaos}.

There is strong evidence that a generic general relativistic system fails to produce Dirac observables and thereby constitutes a weakly non-integrable or even chaotic system. (For the field theory case, the absence of Dirac observables will be a sufficient definition of weak non-integrability too.) The precise definition of chaos for general relativity is unimportant for our purposes. However, we note that there exist appropriate definitions of chaos for generally covariant systems in terms of defocusing and ergodicity of classical trajectories which does not require the notion of exponential divergence \cite{dyncosmo} or in terms of the topological entropy which measures the complexity of the solution space through the number of closed orbits \cite{ottchaos,gutzwillerbook,Cornish:1997ah,Cornish:1996yg,Cornish:1996hx,Motter:2000bg}. What typically happens in systems which are chaotic according to such definitions is that a generic classical trajectory explores regions of large dimension in the constraint surface -- in conflict with the notion of weak integrability above.

We shall list {\it some} of the evidence for weak non-integrability of general relativity in increasing strength. 
\begin{enumerate} 
\item A generic dynamical system is chaotic and this becomes more pronounced the more intricate its dynamics \cite{ottchaos,gutzwillerbook,berrychaos}. Simply by that reasoning -- and in light of the complicated general relativistic dynamics -- it would be a surprise if full general relativity was integrable in any of the above senses.
\item The $N\geq3$ body problem in Newtonian gravity is chaotic \cite{arnold2007mathematical}. Given that general relativity reproduces Newtonian gravity in the appropriate limit but is otherwise technically more intricate, it is hardly imaginable that general relativity remedies the situation.
\item Already one of the simplest cosmological models, namely the closed FRW model with minimally coupled massive scalar field, has been shown to be chaotic \cite{Cornish:1997ah,Page:1984qt,Kamenshchik:1998ue} (see also \cite{Hohn:2011us} for a discussion of relational dynamics in this model). There is analytical and numerical evidence that the space of solutions of this model features a fractal structure. This would impede the existence of a reduced phase space. 
\item There is good evidence that the most general homogeneous cosmological model, the Mixmaster (or Bianchi IX) universe, is chaotic \cite{Misner:1969hg,Cornish:1996yg,Cornish:1996hx,Motter:2000bg}.
\item The well-known Belinsky-Khalatnikov-Lifshitz conjecture \cite{Belinsky:1970ew} posits that a generic (inhomogeneous) cosmological solution to the Einstein equations undergoes chaotic oscillations on approach to the initial singularity.
\item It follows from a classification of generalized symmetries of the vacuum Einstein equations that, for closed spatial manifolds, no Dirac observables exist which are spatial integrals over local functions of the metric and its derivatives \cite{Torre:1993jm,Anderson:1994eg}. This suggests that the vacuum Einstein equations are non-integrable.
\end{enumerate}
It may thus be no surprise that no Dirac observables are known for full general relativity -- simply because they almost certainly do not exist in general in the first place. One should not get fooled and draw erroneous conclusions about gravitational Dirac observables and relational dynamics from our understanding of general relativity through simple integrable systems.

Two important comments are in place. Firstly, this clearly does not imply that no general relativistic systems with Dirac observables exist. As mentioned in the introduction, general relativity subject to special asymptotic symmetries or special matter contents gives rise to Dirac observables. Furthermore, the simplest cosmological models are certainly integrable. But these systems do not account for a generic situation in general relativity. Secondly, weak non-integrability also does not entail the absence of gauge invariant observables. 

{
There are two notions in which gauge invariant observables appear: Firstly as non--differentiable or even non--continuous functions on the constraint surface (see also \cite{Hajicek:1995en,Hajicek:1996xk} for a brief discussion). This will be illustrated by means of toy models in sections \ref{sec_ex1} and \ref{sec_furtherex} below. Thus, there is classically no conceptual problem, gauge invariant quantities exist by means of which one may interpret the dynamics. Yet, such generalized non-differentiable observables are not Dirac observables as they cannot be inserted into Poisson brackets and will thus fail to satisfy Poisson algebraic relations. 

Secondly, it is always possible (around a regular phase space point on the constraint hypersurface) to find sufficiently many smooth gauge invariant functions locally in phase space. That is, these functions will in general fail to be globally well defined or globally invariant. Again this will be illustrated in the examples in the main text. Indeed, the generalized Darboux theorem for constrained systems (see for instance \cite{Tyutinbook}) guarantees that locally there always exist sufficiently many phase space functions that commute with the constraints.

Thus the issue we are discussing here is a highly non--perturbative one and results from global features of the phase space and constraint hypersurface. The drastic effect on quantization should not be a surprise as quantization depends strongly on global features \cite{isham2}. This is again illustrated by the example, e.g.\ in section \ref{sec_1}, which differs from a fully integrable system by a periodic identification in the configuration space.  
}

The reader might wonder why we put much emphasis on general relativity and why weak non-integrability would not be an issue, e.g., for Yang-Mills theories. Certainly, depending on its constraint structure, a general constrained system will feature the same non-integrability issues. What is special about the canonical constraints in Yang-Mills-theories and the diffeomorphism (and even Gauss) constraints in general relativity is that they are linear in the momentum variables (and their derivatives). However, linear constraints usually generate much simpler and integrable dynamics. The trouble maker for general relativity is the Hamiltonian constraint which is quadratic in the momentum variables and generates a much more complicated dynamics. Whence our focus on general relativistic systems.

Lastly, we stress an essential conceptual and technical difference between unconstrained non-integrable systems and totally constrained, weakly non-integrable systems. The absence of differentiable constants of motion in the former does not pose a conceptual challenge: While constants of motion are convenient for fully solving the equations of motion, they are not strictly required for the interpretation of the system because one does not need to solve the dynamics in order to access the physical degrees of freedom. This also holds in the quantum theory where one does not need so solve the Schr\"odinger equation in order to access the physical observables; all observables on the Hilbert space are, in principle, `physical'. Whether the Schr\"odinger equation is difficult to solve or not is an unrelated problem. 

By contrast, in a totally constrained system one firstly needs to solve the dynamics in order to access the physical degrees of freedom -- both in the classical and quantum theory. The `constants of motion' of the constraints represent the physical degrees of freedom which are necessary for the interpretation of the system. Thus, the absence of Dirac observables has different repercussions than non-integrability in the unconstrained case. While classically this does not necessarily constitute a conceptual problem because the generalized observables are available, it potentially becomes both a conceptual and technical problem in the quantum theory since, depending on one's method of quantization, one may not be able to construct a quantum representation of these non-differentiable quantities. This shall be one of the focuses of the main body of this article.

\section{Ergodic free particle dynamics}\label{sec_ex1}

In the remainder, we shall attempt to make some of this discussion concrete by studying simple toy models, rather than considering general relativity. Some of these models fail to be weakly integrable by being irreducible, but none of them will be fully weakly non-integrable. The advantage of working with models that are not fully weakly non-integrable is that they can be handled analytically in contrast to most chaotic systems, especially in a gravitational context (e.g., see the conceptual and numerical discussion in \cite{Hohn:2011us}). Yet, by being irreducible, they already illustrate some of the behaviour generic to weakly non-integrable models.

\subsection{Free particle in the plane}

We begin with a trivial example: consider the Lagrangian with constant $E$
\ba
L=\f{m}{2\lambda}(\dot{x}_1^2+\dot{x}_2^2)+\lambda\,E\label{lag}
\ea
as a function of $(x_1,x_2)\in\mathbb{R}^2$ and $\lambda$ as well as their derivatives in `time' $t$. This leads to the momenta
\ba
p_i=\f{\p L}{\p\dot{x}_i}=\f{m \,\dot{x}_i}{\lambda},\q\q\q\q\pi_\lambda=\f{\p L}{\p \dot{\lambda}}=0\nn,
\ea
where the second equation is a primary constraint. Using the Legendre transformations, the Lagrangian takes the form
\ba
L=\dot{x}_1p_1+\dot{x}_2p_2-\lambda\left(\f{p_1^2}{2m}+\f{p^2_2}{2m}-E\right)
\ea
and immediately leads to a secondary constraint (equation of motion for $\lambda$)
\ba
C=\f{p_1^2}{2m}+\f{p^2_2}{2m}-E\approx0,\label{plane0}
\ea
where $\approx$ denotes a weak equality. That is, $\lambda$ can be an arbitrary function of $t$. Clearly, $C$ and $\pi_\lambda$ commute such that both are first class. Since the dynamics of $\lambda$ is trivial, we disregard it as a dynamical variable and simply treat it as a Lagrange multiplier -- in analogy to the lapse in general relativity. This system thus has the interpretation of a free particle in $\mathbb{R}^2$ with fixed energy. Given the arbitrariness of $\lambda$ as a function of $t$, it is clear that (\ref{plane0}) generates reparametrizations in $t$ such that $t$ is a gauge parameter and not a proper time variable. 

The system is clearly both strongly and weakly integrable and $p_1$ (or, equivalently, $p_2$) and $L_3=x_1p_2-x_2p_1$ constitute a set of independent Dirac observables parametrizing the reduced phase space. Choosing $x_1$ as the relational `clock' and denoting the values it takes by $\tau\in\mathbb{R}$ yields the relational observable
\ba
x_2(\tau)=\frac{p_2}{p_1}\left(\tau-x_1\right)+x_2={\rm{sgn}}(p_2)\f{\sqrt{2mE-p_1^2}}{p_1}\,\tau-\f{L_3}{p_1}\label{relation}
\ea
which, evidently, is invariant under (\ref{plane}). We shall now study how the integrability is affected by compactifying the dynamics on a torus and adding a few modifications to the constraint.

\subsection{Compactification and anisotropic masses}\label{sec_1}

Next, we compactify the dynamics on a torus by identifying $x_i+1\sim x_i$, $i=1,2$ and, moreover, introduce anisotropic masses such that the constraint becomes
\ba
C=\f{p_1^2}{2m_1}+\f{p^2_2}{2m_2}-E\approx0.\label{plane}
\ea
This system can also be viewed as two free particles on a circle and follows from a Lagrangian analogous to (\ref{lag}). The constraint (\ref{plane}) again generates reparametrizations in $t$.

\begin{figure}[hbt!]
\begin{center}
{\includegraphics[scale=1]{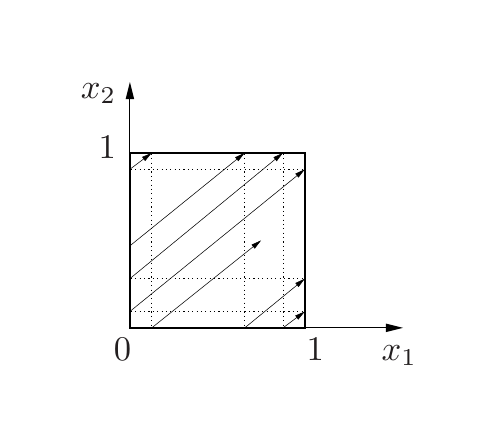}}
\caption{\small A trajectory on the torus $\mathbb{T}^2$ with angle given by (\ref{phi}). $n_i$ increases by $1$ whenever $x_i=1$ is reached.}\label{fig_1}
\end{center}
\end{figure}

The constraint surface corresponding to (\ref{plane}) is compact and has topology
\ba
\cc\simeq I\times\mathbb{T}^2,
\ea
where $I=[-\sqrt{2m_1E},+\sqrt{2m_1E}]$ is a closed interval parametrized by $p_1$ (it could analogously be parametrized by $p_2$). The solutions to the equations of motion generated by (\ref{plane}) read
\ba
x_1(t)&=&\left(\f{p_1}{m_1}t+x_{10}\right)\,\,\text{mod} \,\,1\,\,=\,\,\f{p_1}{m_1}t+x_{10}-\Big\lfloor\f{p_1}{m_1}t+x_{10}\Big\rfloor,\label{bla0}\\
x_2(t)&=&\left(\f{p_2}{m_2}t+x_{20}\right)\,\,\text{mod} \,\,1\,\,=\,\,\f{p_2}{m_2}t+x_{20}-\Big\lfloor\f{p_2}{m_2}t+x_{20}\Big\rfloor,\label{bla}
\ea
where $t$ is the gauge parameter along the flow of $C$, and $\lfloor \cdot\rfloor$ denotes the floor function; here its values $n_i\in\mathbb{Z}$ count the `winding number' in $x_i$. 
For later purpose we note that 
\ba
\tan\phi = \gamma\,\f{p_1}{p_2}\label{phi}
\ea
defines the angle of the trajectory on the flat torus, where $\gamma=m_2/m_1$ is the mass ratio (see figure \ref{fig_1}).

Clearly, as before $p_1$ (or $p_2$) is a valid Dirac observable, parametrizing $I$ and being constant along the trajectories. Since $p_1$ strongly commutes with $C$, the model is strongly integrable and not fully weakly non-integrable. But does there exist another Dirac observable, $F$, in analogy to $L_3$ above, which is independent of $p_1$ and parametrizes the trajectories on $\mathbb{T}^2$ where the non-trivial part of dynamics takes place? $F(p_1;x_1,x_2)$ must satisfy at least one of $\p_1F\neq0$ or $\p_2F\neq0$ and, ideally, 
\ba
\{F,C\}=\f{p_1}{m_1}\p_1F+\f{p_2}{m_2}\p_2F\approx0\label{F}.
\ea

We shall show that no continuous function $F$ on $\cc$ exists which is constant along all trajectories and independent of $p_1$. In particular, $F$ satisfying (\ref{F}) cannot exist. The model thus fails to be weakly integrable. To this end, we must distinguish two qualitatively very different cases:
\begin{description}
\item[resonant tori:]  $\gamma\f{p_1}{p_2}\in\mathbb{Q}$ such that all trajectories on $\mathbb{T}^2$ are closed and periodic \cite{gutzwillerbook,berrychaos,arnold2007mathematical}.
\item[non-resonant tori:] $\gamma\f{p_1}{p_2}\notin\mathbb{Q}$ and the torus is filled densely by ergodic, non-periodic trajectories \cite{gutzwillerbook,berrychaos,arnold2007mathematical}.
\end{description}
Both the resonant and non-resonant tori fill the constraint surface densely since both the rational and irrational numbers fill $I$ densely. However, the rationals have a zero Lebesgue measure in $I$ such that the resonant tori are of Lebesgue measure zero in $\cc$. By contrast, the non-resonant tori are of Lebesgue measure $1$ in $\cc$.

It is the non-resonant tori on $\cc$ which destroy the integrability of the system and the existence of a continuous $F$, independent of $p_1$ and constant along all trajectories. Namely, suppose $\gamma\f{p_1}{p_2}\notin\mathbb{Q}$. Then all corresponding trajectories fill the torus densely, i.e., every neighbourhood of every point on the torus is intersected by each trajectory. Hence, $F$, for this fixed value of $p_1$ has to take all its possible values it may assume on the entire torus in every of its open neighbourhoods. But this violates the definition of continuity. In fact, this situation holds for every point on the torus such that $F$ is nowhere continuous on the torus. This models is therefore not weakly integrable.

In particular, $F$ fails to be differentiable on $\cc$ such that one cannot insert it into Poisson brackets (globally on $\cc$). Accordingly, there does not exist a sufficient Poisson algebra of independent Dirac observables parametrizing the solutions. Consequently, a reduced phase space does not exist. This immediately rules out quantization by the reduction method since we do not have Poisson algebraic structures available which we could promote into a commutator algebra in the quantum theory.

Even more severely, the space of solutions (which for integrable systems constitutes the reduced phase space) is neither Hausdorff nor a manifold. The space of solutions is a quotient space $\cs=\cc/\sim$, where $\sim$ corresponds to the equivalence relation $y\sim y'$ $\Leftrightarrow$ $\alpha_C^t(y)=y'$ for some $t$, where $y,y'\in\cc$ and $\alpha_C^t$ denotes the flow generated by (\ref{plane}):
\begin{itemize}
\item[(i)] \underline{$\cs$ is non-Hausdorff:} The equivalence relation $\sim$ defines a quotient map $Q:\cc\rightarrow \cs$. The open sets $U_S$ in $\cs$ are those which have an open set $U_C\subset\cc$ as pre-image under $Q$. Denote by $\tau_{p_1}$ a trajectory in the torus labeled by $p_1$. Let $p_1$ give rise to a non-resonant torus. Since every $\tau_{p_1}$ densely fills this torus, every open set in $\cc$ which is intersected by a given $\tau_{p_1}$ is also intersected by every other $\tau_{p_1}'$ for the same $p_1$. Indeed, there exist distinct $\tau_{p_1}\neq\tau_{p_1}'$: for example, $(x_1,x_2,p_1,p_2)$ and $(x_1+\f{1}{2},x_2,p_1,p_2)$ with $\gamma\f{p_1}{p_2}\notin\mathbb{Q}$ do not lie on the same trajectory since no winding number $n_1$ exists which maps these two points into one another. Hence, every open set $U_S\subset\cs$ containing $\tau_{p_1}$ also contains every other $\tau'_{p_1}$ such that solutions on a non-resonant torus cannot be distinguished by open neighbourhoods in $\cs$.

\item[(ii)] \underline{$\cs$ is not a manifold:} If $\cs$ was a topological manifold, there would exist a homeomorphism $f_U:U_S\rightarrow V\subset\mathbb{R}^n$ open for all open $U_S\subset\cs$. In fact, $n>1$, by degree of freedom counting (we have $\cs\simeq I\times\ct$ where $I$ is a one-dimensional manifold and $\ct$ denotes the set of trajectories on a torus for fixed $p_1$). Denote by $U_C\subset\cc$ the pre-image under $Q$ of $U_S$, i.e., $U_S=Q(U_C)$. Since $Q$ is continuous by construction, we then have that $f_U\circ Q:U_C\rightarrow\mathbb{R}^n$ is continuous iff $f_U$ is continuous. Notice that $f_U\circ Q$ is constant along all trajectories in $U_C$. However, it follows from the above arguments, that there exist neighbourhoods $\tilde{U}_C\subset\cc$ (intersecting non-resonant tori) such that no continuous $f_{\tilde{U}}\circ Q:\tilde{U}_C\rightarrow\mathbb{R}^{n>1}$ exists which is constant along all trajectories in $\tilde{U}_C$.\footnote{This would require at least two independent continuous functions $\tilde{U}_C\rightarrow\mathbb{R}$ which are constant along trajectories in $\tilde{U}_C$.} But then for $\tilde{U}_S=Q(\tilde{U}_C)$ no homeomorphism $f_{\tilde{U}}:\tilde{U}\rightarrow \mathbb{R}^{n>1}$ exists such that $\cs$ is not a manifold.

\end{itemize}

Therefore, the Marsden-Weinstein symplectic reduction fails in this toy model. 

It should be emphasized that this model is even fairly benign: (1) it is `semi-integrable' since $p_1$ is a Dirac observable, and (2) despite featuring ergodicity in the configuration manifold, the dynamics is {\it not} chaotic (trajectories either always stay parallel or have a linear relative distance growth in the gauge parameter).

%
%
%


For a fully non-integrable -- or even chaotic -- system (and this is the generic case!), the situation will be worse: no differentiable (global) Dirac observables whatsoever will arise, impeding the existence of a reduced phase space and presumably any manifold structure for the space of solutions. Indeed, there is both analytical and numerical evidence that the space of solutions of the chaotic closed FRW model with minimally coupled massive scalar field features a fractal structure \cite{Page:1984qt,Cornish:1997ah}.

Nevertheless, as a generalization of Dirac observables, we can still have gauge invariant `observables': gauge invariance means constance along the orbits generated by the first class constraints and does not require differentiability. As seen above, in a generic case, such `observables' will be highly discontinuous functions without (Poisson) algebraic relations. {(See section \ref{classi} for the construction of such an observable.)} But at the classical level no {\it conceptual} difficulty arises; physical information need not be continuous.

Alternatively, instead of global discontinuous observables, one may construct transient observables \cite{Bojowald:2010xp,Bojowald:2010qw,Hohn:2011us} where temporal locality means validity only for sections of a trajectory. For instance, we can ask for the correlation of the `evolving' $x_2$ with the `clock' $x_1$, in analogy to (\ref{relation}), despite this relation not constituting a global smooth relational Dirac observable due to the above arguments. To this end, it will be convenient to express the winding numbers of the trajectory, $n_i(t):=\lfloor \f{p_i}{m_i}t+x_{i0}\rfloor$ in (\ref{bla0}, \ref{bla}), as functions of $x_1(t),x_2(t)$, rather than $t$. We have to distinguish again between the resonant and non-resonant case.

For a resonant torus, $n_i$ cannot be parametrized uniquely by $x_1(t),x_2(t)$ (and initial data $x_{i0}$) for the entire evolution because of periodicity. However, we can parametrize the $n_i$ for every period of the orbit by $x_1(t),x_2(t)$ and $x_{10},x_{20}$. Let $N_2$ be the number of complete revolutions in $x_2$ before a trajectory closes in on itself. Accordingly, since any fixed $x_1=\tau$ will be passed $N_2$ times during every period, $x_2(x_1=\tau)$ will have $N_2$ solutions. Thanks to periodicity, these will also globally be all possible solutions.

The situation is different for non-resonant tori. In this case, given initial data $x_{10},x_{20}$, $x_1(t),x_2(t)$ uniquely specify the winding numbers for the entire evolution, since the trajectories do not pass the same point twice.\footnote{More precisely, given $x_{10},x_{20}$, $x_1(t),x_2(t)$, the two winding numbers $n_i$ are found as follows: find $t\in\mathbb{R}$ such that
\ba
x_1(t)-x_{10}&=&\f{p_1}{m_1}t - n_1,\nn\\
x_2(t)-x_{20}&=&\f{p_2}{m_2}t - n_2\nn
\ea
gives $n_1,n_2\in\mathbb{Z}$. There is at most one solution. Namely, suppose there exist distinct $t'\in\mathbb{R}$ and $n_1',n_2'\in\mathbb{Z}$ also satisfying the last equations. Then one can easily check that we must have
\ba
\gamma\,\f{p_1}{p_2}=\f{n_1'-n_1}{n_2'-n_2}.\nn
\ea
However, for a non-resonant torus $\gamma\f{p_1}{p_2}\notin\mathbb{Q}$ which cannot be satisfied by $n_i,n_i'\in\mathbb{Z}$ such that $n_1',n_2'$ cannot exist. Clearly, the case of one/no solution corresponds to $x_1(t),x_2(t)$ lying/not lying on the trajectory with initial data $x_{10},x_{20}$.}
 Thus, we may write $n_i(x_1(t),x_2(t);x_{10},x_{20})$ which is implicitly well-defined for pairs $(x_1(t),x_2(t))$ lying on a trajectory defined by $x_{10},x_{20}$. A numerical evaluation, however, would require infinite precision because every trajectory comes arbitrarily close to every point on the torus. There are infinitely many solutions for $x_2(x_1=\tau)$: since every trajectory fills the torus densely, it will also densely cross a loop of fixed $x_1=\tau$. The same applies to any closed loop on the torus such that the Gribov problem of fixing gauge becomes untractable in this model; for every gauge fixing, represented by such a loop, there exist densely many solutions.

\begin{figure}[hbt!]
\begin{center}
{\includegraphics[scale=1]{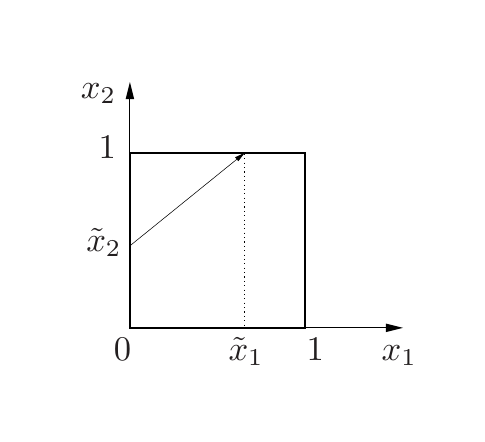}}
\caption{\small Each branch of a trajectory, given by an admissible pair $(n_1,n_2)$, features a linear relation between $x_2$ and $x_1$. The range of permissible values of $x_i$ depends on the branch; in the figure $x_1\in[0,\tilde{x}_1)$ and $x_2\in[\tilde{x}_2,1)$. }\label{fig_2}
\end{center}
\end{figure}

In either case,\footnote{In the resonant case, for a given period.} we can then solve the first equation in (\ref{bla}) for $t$ and insert it into the second equation, while replacing the floor functions by the corresponding $n_i$, to obtain the implicit relation
\ba
x_2(\tau)=\f{1}{\gamma}\frac{p_2}{p_1}\left(\tau-x_{1}+n_1(\tau,x_2(\tau);x_1,x_2)\right)+x_{2}-n_2(\tau,x_2(\tau);x_1,x_2),\label{bla2}
\ea
where $\tau\in[0,1)$ is the value of the `clock' $x_1$, $x_2(\tau)$ is the relational observable yielding the value of $x_2$ when $x_1$ reads $\tau$, and $x_1,x_2$ are the corresponding phase space variables. This relation between the `evolving' $x_2$ and the `clock' $x_1$ is globally well-defined -- albeit implicitly and multi-valued. While it is impossible to globally solve (\ref{bla2}) for $x_2(\tau)$ (there exist `densely many' solutions in the non-resonant case), locally along the trajectory, one can solve (\ref{bla2}) explicitly: for every admissible pair of winding numbers $(n_1,n_2)$, i.e.\ on every `branch' of the solution, we find a linear relation between $x_2$ and $x_1$  (see figure \ref{fig_2}).\footnote{Qualitatively, this is a generalization to continuous dynamics of the branch dependent relational observables of the chaotic discrete time Baker map \cite{Dittrich:2004cb}.} For instance, on the `initial' branch of the evolution we have $n_i=0$ such that, in analogy to (\ref{relation}),
\ba
x_2(\tau)=\f{1}{\gamma}\frac{p_2}{p_1}\left(\tau-x_{1}\right)+x_{2}.\label{localobs}
\ea
We emphasize the temporal locality of these explicit relations; e.g., for an ergodic orbit, this relation is only valid on the `initial' branch as no pair $(x_1,x_2)$ will be passed twice in the evolution. Furthermore, the range of permissible values of $x_2(\tau)$ and $\tau$ is not the full $\mathbb{T}^2$ and depends on the branch $(n_1,n_2)$ (see figure \ref{fig_2}).

Using $n_i(t)=\lfloor \f{p_i}{m_i}t+x_{i0}\rfloor$, it is not difficult to determine all values of $n_2$ compatible with a fixed value of $n_1$ -- and vice versa. In fact, for a trajectory on a resonant torus one can in this manner determine the finitely many admissible pairs of $(n_1,n_2)$ for one period which, in principle, permits to explicitly compute all $N_2$ distinct solutions to (\ref{bla2}). By contrast, for a trajectory on a non-resonant torus there exist infinitely many admissible pairs of $(n_1,n_2)$, each giving rise to a distinct relation between $x_2$ and $x_1$, thereby preventing a global solution to (\ref{bla2}). Nevertheless, locally we have explicit relations and predictions.

\section{Standard quantization}\label{sec_standard}

A reduced phase space and a Poisson algebra of Dirac observables which could be promoted to a commutator algebra do not exist such that quantization by the reduction method is outright impossible. However, remarkably, Dirac quantization \emph{is} possible. This model thus constitutes an extreme example for showing that reduced and Dirac quantization are in general not equivalent. Nevertheless, as we shall see shortly, also this standard Dirac quantum theory is ill and generically devoid of a semiclassical limit. In section \ref{sec_discrete} below, we shall discuss an alternative quantization which remedies this illness.

\subsection{Dirac quantization and physical Hilbert spaces}

We choose wave functions $\psi(x_1,x_2)$ in the kinematical Hilbert space $\ch_{\rm kin}=L^2(S^1 \times S^1)$ with $x_1,x_2 \in [0,1)$. The momenta act as
\ba
\hat p_i \psi &=& -i \hbar \partial_i \psi .
\ea
The momentum eigenstates
\ba
\psi_{k_1,k_2}(x_1,x_2) = \exp(2\pi\, i k_1 x_1)  \exp(2\pi\, i k_2 x_2)   \label{basis}
\ea
with $(k_1,k_2)\in \mathbb{Z}^2$ are a dense orthonormal basis set in $\ch_{\rm kin}$, diagonalizing the constraint
\ba
\hat{C}=\f{\hat{p}_1^2}{2m_1}+\f{\hat{p}^2_2}{2m_2}-E\nn
\ea
with discrete spectrum
\ba
\Big\{  2(\pi\hbar)^2 \left(\f{k_1^2}{m_1} +\f{ k_2^2}{m_2}\right) - E \,|\, (k_1,k_2)\in \mathbb{Z}^2 \Big\}  .\label{spectrum}
\ea

Solutions to the constraint $\hat{C}\,\psi_{\rm phys}=0$ are defined by $(k_1,k_2)\in\mathbb{Z}^2$ such that
\ba
k_1^2+\f{1}{\gamma}k_2^2=\epsilon:=\f{m_1E}{2(\pi\hbar)^2}.\label{equ:constraintSpectrum}
\ea
Solutions exist only for special (non-generic) $E \in \mathbb{R}_+$ for which $0$ will be in the spectrum.  

It is also clear that if $\psi_{k_1,k_2}$ is in the constraint kernel then so are $\psi_{k_1,-k_2},\psi_{-k_1,k_2},\psi_{-k_1,-k_2}$. Moreover, if $\psi_{k_1,k_2}$ is in the constraint kernel then $\psi_{k'_1,k'_2}$ can only be in the constraint kernel if
\begin{equation}
 \gamma=\frac{k^2_2-k'^2_2}{k'^2_1-k_1^2},\label{gamma}
\end{equation}
where $\infty>\gamma>0$. It follows that we have to distinguish the case $\gamma\notin\mathbb{Q}$ from $\gamma\in\mathbb{Q}$:

For $\gamma\notin\mathbb{Q}$, there are only four cases:
\begin{enumerate}
 \item $\dim(\ker(\hat{C}))=0$, the generic $\epsilon$ for which no pair of integers $k_i$ exist that solves (\ref{equ:constraintSpectrum});
 \item $\dim(\ker(\hat{C}))=1$, for the special case $\epsilon=0$, where both $k_i$ vanish;
 \item $\dim(\ker(\hat{C}))=2$, for special values of $\epsilon=k^2$ or $\epsilon=\f{1}{\gamma}\,k^2$, when one $k_i$ vanishes and the other equals $\pm k$;
 \item $\dim(\ker(\hat{C}))=4$, for the special values $\epsilon=k_1^2+\f{1}{\gamma}\,k_2^2$.
\end{enumerate}
Higher dimensional kernels do not arise since (\ref{gamma}) cannot be satisfied for $\gamma\notin\mathbb{Q}$ and $k_i,k_i'\in\mathbb{Z}$.

By contrast, the case of rational $\gamma$ is an intricate Diophantine problem, which possesses isolated values $\epsilon$ with many solutions, e.g.\ for $\gamma=\frac 1 2$ one can solve $\epsilon=9$ as $(k_1,k_2)=(3,0)$ or as $(k_1,k_2)=(1,4)$. 

We illustrate the discussion for the equal mass case $\gamma=1$, where the problem is related to a well understood problem in number theory: In how many distinct ways $w(n)$ can a given natural number $n$ be written as the sum of squares of two integers $k_1,k_2$? We have to distinguish two cases, since $w(n)$ regards $k_1$ and $k_2$ as unordered, while in the constraint kernel $k_1$ and $k_2$ have to be considered as ordered to appropriately count the dimension. Hence, if there is no integer $k$ such that $n = 2 k^2$ then the dimension of the kernel is $2\,w(n)$, otherwise it is $2(w(n)-1)$. A classical result by Jacobi \cite{Jacobi1829:aa} states that $w(n)$ for a given integer 
\begin{equation}
 n=2^a\,p_1^{b_1}\cdots\,p_n^{b_n}\,q_1^{c_1}\cdots\,q_m^{c_m},
\end{equation}
where the $p_i$ are the prime factors of $n$ with $p_i\, \textrm{mod} \,\,4=3$ and the $q_i$ are the prime factors of $n$ with $q_i\, \textrm{mod}\,\, 4=1$, vanishes if at least one of the $b_i$ is odd (which is the bulk of integers $n$), and, if all $b_i$ are even, the number is
\begin{equation}
 w(n)=2(c_1+1)\cdots(c_m+1).
\end{equation}
We see that the number of states in the constraint kernel is not bounded from above: There are always exceptional values of $\epsilon$, such as $\epsilon=5^{d }$ for $d\in\mathbb N$, for which the constraint kernel has dimension $4(d+1)$. For example, for $\epsilon=25=5^2$ one has $4(2+1)=12$ states in the constraint kernel labeled by momentum eigenvalues $(k_1,k_2)=$ $(0,-5),\,$ $(0,5),\,$ $(-5,0),\,$ $(5,0),\,$ $(-3,-4),\,$ $(3,-4),\,$ $(-3,4),\,$ $(3,4),\,$ $(-4,-3),\,$ $(4,-3),\,$ $(-4,3),\,$ $(4,3)$; or for $\epsilon=71825=5^2\,13^2\,17$ one has $4(2+1)(2+1)(1+1)=72$ states in the constraint kernel labeled by the momentum eigenvalues $(\pm 15, \pm 400)$, $(\pm 32, \pm 399)$, $(\pm 76, \pm 393)$, $(\pm 81, \pm 392), (\pm 113,  \pm 384)$, $(\pm 140, \pm 375)$, $(\pm 175, \pm 360)$, $(\pm 183,\pm 356)$, $(\pm 216,\pm 33)$, $(\pm 228, \pm 329)$,\newline $(\pm 252, \pm 311)$, $(\pm 265, \pm 300)$. In conclusion, large constraint kernels exist but are rather exotic. 

In either case, the constraint kernel -- and thus the physical Hilbert space $\ch_{\rm phys}$ -- is always finite dimensional. $\ch_{\rm phys}$ is spanned by the momentum eigenstates satisfying (\ref{equ:constraintSpectrum}) and, since the spectrum (\ref{spectrum}) is discrete, a proper subspace of the kinematical Hilbert space $L^2(S^1\times S^1)$, thereby inheriting its inner product.

\subsection{Physical observables}

Being finite dimensional, the physical Hilbert spaces are isomorphic to spin Hilbert spaces with observables on them taking the form of spin observables. In particular, no canonically conjugate pairs of Dirac observables exist on $\ch_{\rm phys}$. For example, the largest possible physical Hilbert spaces for $\gamma\notin\mathbb Q$ are isomorphic to a two-qubit Hilbert space $\ch_{\rm phys}\simeq\mathbb{C}^4$. This is a very small Hilbert space with few observables; a linearly independent basis set of observables on $\mathbb{C}^4$ is comprised of 15 binary propositions and the identity (e.g., see \cite{Hoehn:2014uua} for a formulation in terms of questions). It can already be anticipated to be impossible to equip quantum theories on such Hilbert spaces with relational observables admitting an interpretation similar to that in the classical theory.

Clearly, the momenta $\hat{p}_i$ are observables on $\ch_{\rm phys}$, but also take the form of spin observables. The momentum eigenstates in $\ch_{\rm phys}$ correspond to resonant tori for $\gamma\in\mathbb Q$ and to non-resonant tori for $\gamma\notin\mathbb{Q}$ since the ratio of the momentum eigenvalues is always rational. We shall exploit this shortly. However, we emphasize that, by means of linear combinations, one can engineer $\gamma\f{\langle\hat{p}_1\rangle_{\rm phys}}{\langle\hat{p}_2\rangle_{\rm phys}}$ to be either rational or irrational, where $\langle\cdot,\cdot\rangle_{\rm phys}$ is the inner product on $\ch_{\rm kin}$ restricted to the set of solutions to the constraint.\footnote{For instance, suppose $\gamma=1$ and $\epsilon=2k^2$. The physical state
\ba
\psi=\alpha\,\exp(2\pi\,i\,k(x_1+x_2))+\beta\exp(2\pi\,i\,k(x_1-x_2))\nn,
\ea
can be easily checked to yield a ratio
\ba
\f{\langle\hat{p}_1\rangle_{\rm phys}}{\langle\hat{p}_2\rangle_{\rm phys}}=\f{1}{|\alpha|^2-|\beta|^2}\nn
\ea
which, depending on $\alpha,\beta\in\mathbb{C}$, can fail to be in $\mathbb{Q}$.}

What about observables independent of the momenta? Classically, we had either global discontinuous or transient observables taking the place of Dirac observables involving positions. Unfortunately, the global discontinuous observables cannot be represented on $\ch_{\rm phys}$ due to the discontinuity, while local observables of the type (\ref{localobs}) cannot be represented as self-adjoint operators on $\ch_{\rm phys}$ since they are not even classically globally defined. Leaving the issue of self-adjointness aside, one may hope that one can nevertheless give the local observables meaning -- at least in the semiclassical limit as in \cite{Bojowald:2010xp,Bojowald:2010qw,Hohn:2011us,Rovelli:1990jm,Rovelli:1989jn}. However, we shall now argue that this Dirac quantization of the toy model generically does not admit a semiclassical limit. Accordingly, relational dynamics cannot be meaningfully formulated; in particular, a relation to the classical relational dynamics cannot be established.

Qualitatively, we seem to face the following issue: {\it if there are few classical Dirac observables which can be nicely represented in the quantum theory, we have few physical states which can be distinguished by these observables.} We shall circumnavigate this issue with a `polymer' type quantization which admits the representation of more observables in section \ref{sec_discrete}.

\subsection{A quantum illness: Absence of semiclassical states}

The generic failure of a suitable semiclassical limit is an immediate consequence of the previous discussion. Ideally, a proper semiclassical state should (a) be peaked on canonically conjugate Dirac observables, and (b) saturate the uncertainties in these Dirac observables. The trouble is, of course, that we do not even have canonically conjugate Dirac observables such that we cannot obtain such a semiclassical state in the first place. 

Notwithstanding, a finite dimensional constraint kernel and thereby an absence of canonically conjugate Dirac observables on $\ch_{\rm phys}$ does not necessarily inhibit the construction of approximately semiclassical coherent states. For example, in the 2D oscillator model with fixed energy discussed in \cite{Rovelli:1990jm,Bojowald:2010qw} the physical Hilbert spaces are likewise finite dimensional. However, there is a qualitative difference between that model and the present one: in the 2D oscillator model of \cite{Rovelli:1990jm,Bojowald:2010qw} one has a reduced phase space and sufficiently many Dirac observables, one of which is analogous to an angular momentum with dependence on both position and momenta. These Dirac observables are even conjugate on the kinematical Hilbert space of which the physical Hilbert space is a proper subspace. One can peak on these Dirac observables, using the techniques in \cite{ell}, to produce sufficiently semiclassical elliptic coherent states for sufficiently large energies \cite{Bojowald:2010qw}. By contrast, in the present model, we do not have any Dirac observables involving the positions and generically there simply are too few physical states in order to build semiclassical wave packets. For such spin type Hilbert spaces, one would usually constructs semiclassical states using spin coherent states which, however, are not appropriate in the present case.

Nevertheless, let us investigate what can be achieved by constructing `wave packets'. In general, one could imagine two distinct notions of semiclassical wave packets:
\begin{enumerate}
 \item {\it local wave packets:} these are wave packets that are highly peaked near a (small) segment of a classical orbit, but which are allowed to decohere after finite internal time. Such local wave packets would be appropriate for the local relational observables -- if existent.
 \item {\it global coherent states:} these are wave packets that are required to be peaked along an entire classical orbit, which means that these states are not allowed to loose coherence in internal time.  
\end{enumerate}

As before, let us choose $x_1$ as the relational clock. We shall consider the initial wave function at $x_1=0$.\footnote{We are in the special case in which the space of initial wave packets is a tensor factor of $\mathcal H_{\rm kin}$. The space of initial wave functions is thus equipped with the inner product on this tensor factor.} Furthermore, as in \cite{Bojowald:2010qw} we factorize the classical constraint as $C=C_-\,C_+$, where $C_\pm=p_1\pm\sqrt{2m_1E-\f{1}{\gamma}p_2^2}$ generates evolution backward/forward in $x_1$ time. The quantum constraint (\ref{equ:constraintSpectrum}) then implies that the wave function satisfies a Schr\"odinger equation:
\begin{equation}\label{equ:effectiveHamiltonian}
 i\,\hbar\,\frac{\partial\,\psi}{\partial x_1}=\hat{H}\,\psi=\hat{s}\,\sqrt{2m_1E-\f{1}{\gamma}\,\hat{p_2}^2}\,\,\psi,
\end{equation}
where the operator $\hat{s}$ with spectrum $\pm 1$ arises as a quantization of the sign ambiguity of the spectral square root. The classical direction (\ref{phi}) is promoted to an operator
\begin{equation}
  \tan\hat\phi=\gamma\, \hat{p}_1\,\,\widehat{p^{-1}_2}\nn.
\end{equation}
Clearly, the momentum eigenstates are eigenstates of the direction operator. The converse is not necessarily true: e.g., $\alpha\,\psi_{k_1,k_2}+\beta\,\psi_{-k_1,-k_2}$ is an eigenstate of the direction operator but not the momenta. The physical interpretation of the tangent of the direction is the dimensionless speed of the effective evolution. 

We shall now use (\ref{equ:effectiveHamiltonian}) to evolve initial wave functions into physical states resembling wave packets. To this end, we again have to distinguish the cases $\gamma\in\mathbb Q$ and $\gamma\notin\mathbb Q$. We begin with the latter.

~\\
\noindent \underline{$\gamma\notin\mathbb Q$:}  \\[1pt] 

We exclude the case in which one or both $k_i$ vanish, since this implies that the initial wave packet is either completely delocalized or that there is no time evolution. We thus restrict to $\dim\ch_{\rm phys}=4$.
\begin{figure}
 \includegraphics[width=.4\textwidth]{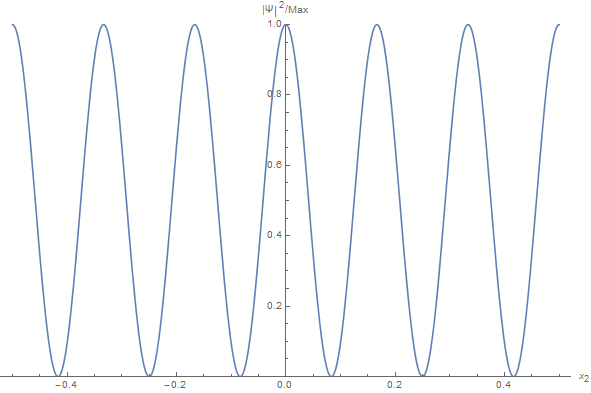}
 \includegraphics[width=.4\textwidth]{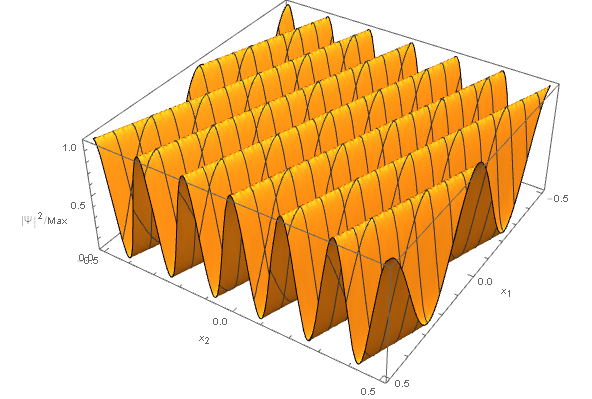}
 \caption{Example quasi-localized wave packet with $k_1=1$, $k_2=3$. 
   \newline {\it Left:} Initial wave function at internal time $x_1=0$. 
   \newline {\it Right:} Global evolution with the right-moving Hamiltonian.}\label{fig:coherentEvolution}
   \end{figure}
The space of initial wave packets at $x_1=0$ is two-dimensional with modes: $\exp(\pm 2\pi i \,k_2\,x_2)$. Thus, the most general initial wave function is of the form
\begin{equation}
 \psi(0,x_2)=\alpha\,e^{2\pi i k_2x_2}+\beta\,e^{-2\pi ik_2x_2}\nn.
\end{equation}
Choosing $\hat{s}=\pm\textrm{sign}(\hat{p_2})$ in (\ref{equ:effectiveHamiltonian}), $\hat{H}$ 
evolves the initial wave function to a solution of the constraint:
\ba
\psi_\pm(x_1,x_2)=\alpha\,e^{2\pi i(k_2x_2\mp k_1x_1)}+\beta\,e^{-2\pi i(k_2x_2\mp k_1x_1)}.\nn
\ea
It can be easily checked that independent of $\alpha,\beta$
\ba
\f{\langle\hat{p}_1\rangle_{\rm phys}}{\langle\hat{p}_2\rangle_{\rm phys}}=\mp\f{k_1}{k_2},\q\q\q\langle\tan\hat{\phi}\rangle_{\rm phys}=\mp\gamma\f{k_1}{k_2}\label{momrat}.
\ea
The associated probability density reads
\ba
|\psi_\pm(x_1,x_2)|^2=|\alpha|^2+|\beta|^2+2\Re[\alpha^*\beta]\,\cos\left(4\pi k_2(x_2\mp\f{k_1}{k_2}x_1)\right)-2\Im[\alpha^*\beta]\,\cos\left(4\pi k_2(x_2\mp\f{k_1}{k_2}x_1)\right)
\ea
It is straightforward to verify that the directions of constant $|\psi_\pm|^2$ on the torus are $(1,\pm\f{k_1}{k_2})$. That is, the evolution is coherent with dimensionless speed $\pm\frac{k_1}{k_2}$ -- independent of $\alpha,\beta$. 

But this coherence is misleading. These states are not semiclassical at all: (1) the localization of these wave packets in position space is too weak to be interpreted semiclassically: The width of a peak in $x_2$ is $1/k_1$ and equals the separation of two subsequent peaks, so the ratio of width to separation is 1. Good localization would require this ratio to be significantly smaller than 1. This does not come as a surprise since there are no position Dirac observables. (2) It can also be checked that the spreads in the momentum Dirac observables are proportional to the energy which is not a semiclassical behaviour. Finally, (3) there is a mismatch in directions. We just saw that both the dimensionless speed and the ratio of the momentum expectation values are $\pm\f{k_1}{k_2}$, while the expectation value of $\tan\hat{\phi}$ differs by $\gamma\notin\mathbb{Q}$. This means that the peak of the `coherent wave packet' does not actually follow a classical trajectory since the latter would have the direction determined by $\langle\tan\hat{\phi}\rangle_{\rm phys}$. The wave packet is peaked on a periodic curve since $k_i\in\mathbb Z$, while a classical trajectory corresponding to these expectation values would be ergodic.

We illustrate these issues in figure \ref{fig:coherentEvolution} for $\alpha=\beta$ which yields the states
\begin{equation}
 \psi_\pm(x_1,x_2)=\alpha\,\cos\left(2\pi \,k_2\left(x_2\mp\frac{k_1}{k_2}x_1\right)\right).\nn
\end{equation}

Since we have exploited a full basis of $\ch_{\rm phys}$, this is as good as it gets for $\gamma\notin\mathbb{Q}$. We thus conclude that in this case the quantum theory is sick and does not possess a semiclassical limit. We also note that this is the generic case because the irrational numbers are of Lebesgue measure 1 (in any finite interval) in the reals.

~\\
\noindent \underline{$\gamma\in\mathbb Q$:}  \\[1pt] 

\begin{figure}
 \includegraphics[width=.4\textwidth]{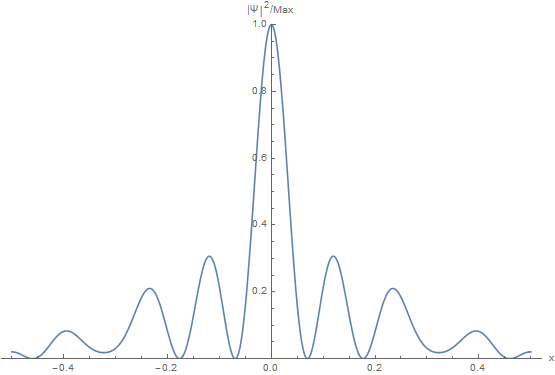}
 \includegraphics[width=.4\textwidth]{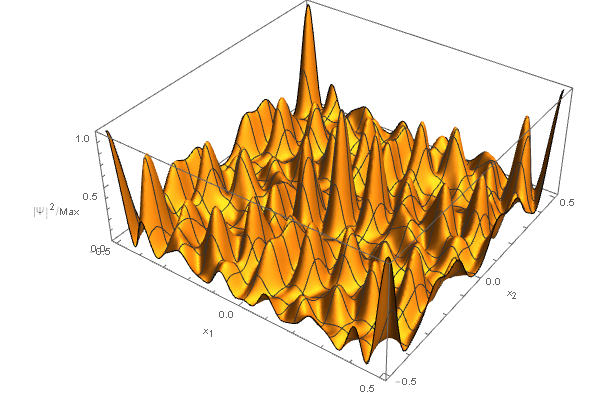}
 \caption{Example of an initially well localized wave packet with mass ratio $\gamma=1$ and $\epsilon=25$.
    \newline {\it Left:} Initial wave function at internal time $x_1=0$. 
   \newline {\it Right:} Global evolution with the right-moving Hamiltonian.}\label{fig:localized25}
   \end{figure}

\begin{figure}
 \includegraphics[width=.4\textwidth]{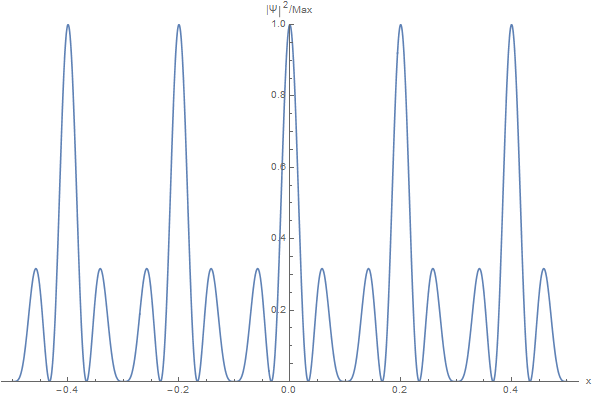}
 \includegraphics[width=.4\textwidth]{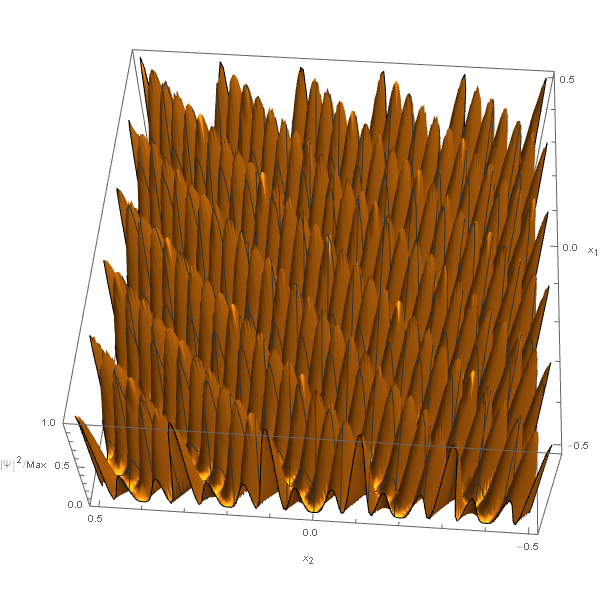}
 \caption{Example of a wave packet with mass ratio $\gamma=1$ and $\epsilon=125$ that is initially localized on an orbit with periodicity 5.
\newline{\it Left:} Initial wave function at initial time $x_1=0$.
\newline{\it Right:} Global evolution with the right-moving Hamiltonian.}\label{fig:coherent25}
   \end{figure}

The discussion for values of $\epsilon$ with small constraint kernels is completely analogous to the irrational case. Differences arise when the constraint kernel is larger. The simplest example with $\gamma=1$ and $\epsilon=25$ resp.\ $\epsilon=125$, which is illustrated in figures \ref{fig:localized25} and \ref{fig:coherent25}. 
The figures show that it is possible to use the allowed $x_2$-frequencies $0,3,4,5$ resp.\ $2,5,10,11$ to build initially quite well localized wave packets, which are either localized at a point (figure \ref{fig:localized25}) or at the intersection of an orbit (figure \ref{fig:coherent25}). The price for the improved localization is decoherence: A coherent wave packet is of the form $\psi(x_2\pm\tan(\phi)\,x_1)$, which would require a linear effective Hamiltonian $\omega(k)=\alpha k$ in order to be generated from a general initial superposition
\ba
\psi(0,x_2)=\sum_i\,\alpha_i\,e^{2\pi i k_2^ix_2}\label{sup}
\ea
where $i$ runs over {\it more} than two values.\footnote{A coherent wave packet is of the form $\sum_k a_k e^{i n_k (x_2+\alpha x_1)}$, from which we read off that coherence requires $\omega(k)=\alpha\,k$.} (For superpositions with only maximally two of the $\alpha_i\neq0$ the same arguments as for $\gamma\notin\mathbb Q$ apply.) Instead, the non-linear Hamiltonian (\ref{equ:effectiveHamiltonian}) that appears in the solution of the constraint (\ref{equ:constraintSpectrum}) yields $\omega(k)=\sqrt{2m_1E-\f{(2\pi\hbar)^2}{\gamma}\,k^2}$ which evolves (\ref{sup}) into non-trivial superpositions
\ba
\psi(x_1,x_2)=\sum_j\,\beta_j\,\psi_j(x_2\pm\tan(\phi_j)\,x_1),\nn
\ea
with distinct $\phi_j$ which does not comprise a coherent wave packet on the torus.\footnote{By contrast, we emphasize that the coherence conditions for the relational evolution of the initial wave packets in the 2D oscillator model of \cite{Rovelli:1990jm,Bojowald:2010qw} is elliptic such that the non-linear square-root Hamiltonian of that model does produce approximately semiclassical states which do permit a temporally local relational dynamics \cite{Bojowald:2010qw}.}

It follows from periodicity of the wave function and the nonlinearity of the effective Hamiltonian, that there can not be any well localized initial states that evolve coherently for long internal time. However, there are conditions for having a good initial localization and short term coherence. Good localization requires many allowed $x_2$-frequencies that are small integer multiples of one another, while short term coherence requires that the Hamiltonian has a good linear approximation over the range of these frequencies. This requires that the allowed frequencies lie in an interval much smaller than $\sqrt{2m_1E}$. However, even such very special wave packets will loose coherence. The better the state is localized initially, the faster it will run apart. That is, global coherent states generically do not exist and local wave packets decohere very quickly. In fact, decoherence will generically happen on scales much smaller than the classical range of validity of a local relational observable of the type (\ref{localobs}). As mentioned before, a quantum version of such local relational observables requires states which are coherent  for sufficiently long. Thus, it is difficult to equip such local quantum observables with a tenable interpretation.

\subsection{Physical transition amplitudes and failure of semiclassicality}\label{section:Tim}

Alternatively, one might try to probe the semiclassical sector of a theory by considering the transition amplitude between kinematical semiclassical sates \cite{Rovelli:2004tv,Rovelli:2001bz} in order to evade the troubles of building semiclassical physical states. But, as we shall now argue, also such transition amplitudes do not admit semiclassical behaviour in the present case. 

Consider two wave functions $|(\vec x_1,\vec p_1)\rangle$ and $|(\vec x_2,\vec p_2)\rangle$, which are kinematical semiclassical wave functions peaked at $(\vec x_i,\vec p_i)$, then we consider
\begin{equation}
 W(\vec x_1,\vec p_1;\vec x_2,\vec p_2)=\frac{\left\langle (\vec x_2,\vec p_2)\right|\hat P\left|(\vec x_1,\vec p_1)\right\rangle}{\sqrt{\left\langle (\vec x_1,\vec p_1)\right|\hat P\left|(\vec x_1,\vec p_1)\right\rangle\left\langle (\vec x_2,\vec p_2)\right|\hat P\left|(\vec x_2,\vec p_2)\right\rangle}},
\end{equation} 
where $\hat P$ denotes the projector onto the kernel of the Hamilton constraint. This is the timeless analogue of the time-dependent transition amplitude $W(\vec x_1,\vec p_1,t_1;\vec x_2,\vec p_2,t_2)=\left\langle (\vec x_2,\vec p_2)\right|\hat U(t_1,t_2)\left|(\vec x_1,\vec p_1)\right\rangle$, where $\hat U(t_1,t_2)$ denotes the time evolution operator. 

The  states are dynamically suitable semi--classical states with respect to the time evolution operator $U$  if (1) the transition amplitude $|W(\vec x_1,\vec p_1,t_1;\vec x_2,\vec p_2,t_2)|^2\approx 1$ whenever the time evolution of $(\vec x_1,\vec p_1)$ from $t_1$ to $t_2$ is $(\vec x_2,\vec p_2)$ and (2) $|W(\vec x_1,\vec p_1,t_1;\vec x_2,\vec p_2,t_2)|^2$ decays rapidly\footnote{E.g., the overlap function of two harmonic oscillator coherent states decays with the exponential of the phase space distance squared.} as $(\vec x_1,\vec p_1)$ moves away from the correct time evolution $(\vec x_2,\vec p_2)$. 

In analogy with this, one says that the   states are dynamically suitable  semi--classical states with respect to the projector $\hat P$  if (1) $|W(\vec x_1,\vec p_1;\vec x_2,\vec p_2)|^2\approx 1$ whenever $(\vec x_1,\vec p_1)$ and $(\vec x_2,\vec p_2)$ lie on the same classical orbit and (2) $|W(\vec x_1,\vec p_1;\vec x_2,\vec p_2)|^2$ decays rapidly when one moves away from an orbit. It is clear that the second requirement is insufficient for ergodic classical orbits, because an ergodic orbit that contains $\vec x_1$ will come arbitrarily close to any $\vec x_2$ and hence $|W|$ can not possess any position dependence. We will see this explicitly at the end of this subsection.

The starting point for the semiclassical discussion is the fixing of a set of semiclassical sates. A particularly versatile set of semiclassical states, which we will use here, are the so-called complexifier coherent states \cite{hall1994segal,hall2002coherent,winkler}. These are a straightforward generalization of the harmonic oscillator coherent states. The starting point for the construction of complexifier coherent states is the choice of an generalized Laplacian $\hat C$, e.g. $\hat C=\frac 1 2 \hat{\vec p}\hat{\vec p}$. 

The operator $\hat C$ is called complexifier, because it is the quantization of a classical phase space function $C$, which is used to complexify the position variables $x_a$ using the definition
\begin{equation}
 z(x_a):=\sum_{n=0}^\infty \frac{i^n}{n!}\{x_a,C\}_{n},
\end{equation} 
where $\{\cdot,C\}_{n}$ denotes the $n$-th nested  Poisson bracket with $C$. E.g. for $C=\frac t 2 \vec p \cdot\vec p$ (with $t$ the so called classicality parameter) one obtains the canonical complex phase space coordinates
\begin{equation}
 z(x_a)=x_a+i t \,p^a.
\end{equation}
The construction of the the complexifier coherent states is then straightforward:
\begin{enumerate}
 \item One writes a formal wave function $\delta_{\vec x_o}(\vec x)$ that is completely localized at a point $\vec x_o$ in configuration space as a formal sum $\delta_{\vec x_o}=\sum_{ k \in B} \overline{\psi_k(\vec x)}|e_k\rangle$ over a basis $B$.
 \item Define $|(\vec x_o)_C\rangle=e^{-\hat C/\hbar}\, \delta_{x_o}:=\sum_{k\in B} \overline{\psi_k(\vec x)}\,e^{-\hat C/\hbar}\,|e_k\rangle$. The sum $|(\vec x_o)_C\rangle$ converges in the Hilbert space for suitable $\hat C$.
 \item Consider $|(\vec x_o)_C\rangle$ as a function of $\vec x_o$ and replace $\vec x_o$ with $z(\vec x_o)$. This defines the coherent state $|(\vec x_o,\vec p_o)_C\rangle:=N\biggl.|(x_o)\rangle\biggr|_{\vec x_o\to z(\vec x_o)}$ for suitable $\hat C$ up to the normalization $N$.
 \item Normalize the states by setting $N=(\langle(\vec x_o,\vec p_o)_C|(\vec x_o,\vec p_o)_C\rangle)^{-\frac 1 2}$.
\end{enumerate}

 For instance the usual harmonic oscillator coherent states can be constructed from $\delta_{x_o}=\int dk \,e^{-ik x_o}|k\rangle$, where $|k\rangle$ denote momentum eigenstates, and the complexifier is given as $\hat C=\frac{t}{ 2} \hat p^2$ with $z(x)=x+itp$. Following the recipe one obtains as complexifier coherent states 
 \ba
 |(x_o,p_o)_{\frac t 2 \hat p^2}\rangle &=& N\int dk \,e^{-\frac t 2 \hbar k^2 - i k x_o + k t p_o }|k \rangle \nonumber\\
 &=&
 N    e^{\frac{t}{2 \hbar}p_o^2}  \int dk \,e^{-\frac{t}{2 \hbar} ( \hbar k -  p_o )^2 - i k x_o  }  |k \rangle
 \ea
 which are indeed the momentum space representation of the harmonic oscillator coherent states.

We are now in the position to construct kinematic coherent states on a unit torus. We choose the standard complexifier $C=\frac t 2 p^2$ and obtain $z(\vec x)=\vec x+it\vec p$. The construction then yields
\begin{equation}
  |(\vec x_o,\vec p_o)_C\rangle =  N    e^{\frac{t}{4\pi \hbar} \vec{p_o}^2 } \sum_{\vec k\in \mathbb Z^2} e^{-\frac{t}{4\pi \hbar } (\vec p_o- 2\pi \hbar\vec k)^2 -  2\pi i\vec k \vec x_o}\,|\vec k\rangle,
\end{equation}
where  $|\vec k\rangle$ denotes  a momentum eigenstate, now with discrete label $\vec{k} \in \mathbb{Z}^2$. 

 
 We can then derive  the physical inner product between these semi--classical states as
\begin{equation}
  \langle(\vec x_1,\vec p_1)|\hat P|(\vec x_2,\vec p_2)\rangle\,= {\cal N}^2 \sum_{\vec k\in \mathbb Z^2:k_1^2+k_2^2/\gamma=\epsilon}
  e^{-\frac{t}{4\pi \hbar} \left((\vec p_1- 2\pi \hbar \vec k)^2+(\vec p_2-2\pi \hbar \vec k)^2\right) -  2\pi i \vec k(\vec x_2-\vec x_1)} 
  \quad .
\end{equation} 
where  in ${\cal N}$ we absorbed the exponential factors $ e^{\frac{t}{4\pi \hbar}\vec{p}_i^2} $. (These factors will eventually drop out in $W$.) Here  we used the projector onto the constraint kernel $\hat P = \sum_{\vec k\in \mathbb Z^2:k_1^2+k_2^2/\gamma=\epsilon} |\vec n\rangle\langle \vec n|$ with $\epsilon=\frac{m_1 E}{2(\pi \hbar)^2}$. 

Any significant contribution to the sum will come from terms in which $\vec p_1$ and $\vec p_2$ are simultaneously close to some $2\pi\hbar\vec k_o$ of the allowed $\vec k$. If $\vec p_1$ and $\vec p_2$ are far apart, then there will be no $\vec k_o$ and $\langle(\vec x_1,\vec p_1)|\hat P|(\vec x_2,\vec p_2)\rangle$ will be exponentially small. However, the two terms $\langle(\vec x_i,\vec p_i)|\hat P|(\vec x_i,\vec p_i)\rangle$ may still be large, if each $\vec p_i$ lies close to an allowed $\vec k_i$. 

In the generic case all summands except for one will be exponentially suppressed because the allowed $\vec k$ do in general not lie close together. In this case the approximation gives 
\begin{equation}
  |W(\vec x_1,\vec p_1;\vec x_2,\vec p_2)| \, \approx \frac{  e^{- \frac{t}{4\pi \hbar} ( (\vec p_1- 2\pi \hbar k_o)^2 + (\vec p_2-2\pi\hbar k_o)^2)}}{e^{- \frac{t}{4\pi \hbar} ((\vec p_1- 2\pi \hbar k_1)^2 + (\vec p_2-2\pi\hbar k_2)^2))  }}.
\end{equation}
This will be exponentially small unless $\vec k_o=\vec k_1=\vec k_2$. In the later case $|W|\approx 1$. This means that $|W|$ encodes information  about whether $\vec p_1$ and $\vec p_2$ lie close to the same allowed value $\vec 2\pi \hbar k_o$ or not. Moreover, we see that $|W|$ has no position dependence in the generic case. This lack of position dependence has a simple interpretation for ergodic orbits: The orbit $(\vec x_1,\vec p_o)$ comes arbitrarily close to any point $\vec x_2$. However, the present discussion applies to periodic orbits as well, as long as there is only one $2\pi\hbar \vec k_o$ that is simultaneously close to $\vec p_1$ and $\vec p_2$. Hence, there is a generic problem with the semiclassical behavior of $W$.

~\\
In conclusion, the standard Dirac quantization of this model is ill because a semiclassical limit generically does not arise. We emphasize that this is qualitatively different from the relation between the classical and quantum theory in unconstrained chaotic systems where ergodic orbits can also not be traced out by semiclassical states \cite{gutzwillerbook,ottchaos}. Firstly, in many chaotic models one obtains so-called `quantum scars' in the semiclassical limit which are increased probability densities along classically unstable periodic orbits, thereby at least approximately tracing out classical periodic orbits --  in contrast to here. Secondly, the Hilbert space is usually infinite dimensional and separable which permits to build initially well-localized wave packets. Such wave packets will also run apart due to the sensitivity to initial conditions but one generally obtains at least a short time coherence which is more stable than in the present case. Thirdly, the conceptual consequences are different in this case: in the unconstrained chaotic case, there is no interpretational issue as one always has a unitary evolution with respect to an absolute time such that the fluctuations and decohering wave packets do not cause any difficulties in interpreting the dynamics. By contrast, in the present case, there is no absolute time and a relational interpretation of the dynamics is generically not possible with strongly decohering wave packets \cite{Bojowald:2010xp,Bojowald:2010qw,Hohn:2011us}. As seen, in the present case, even the more general interpretation in terms of transitions amplitudes is difficult.

We emphasize that this model is still harmless: Part of our analysis was only possible because the momenta are Dirac observables. For instance, the direction operator is only meaningful thanks to these observables. The situation will be much worse for systems entirely deprived of global Dirac observables. The present analysis suggests that it might be impossible to meaningfully Dirac quantize such a system in this standard manner. Indeed, in section \ref{sec_screwedup}, we shall see a model which only permits ergodic orbits classically and cannot be quantized in the standard Dirac manner, although not even being chaotic. In this case, without a quantum theory, the issue of absence of semiclassicality does not even pose itself.

\section{Introducing a discrete topology on the torus}\label{sec_discrete}

As discussed in section \ref{sec_nonint} and illustrated in the toy model in section \ref{sec_1}, the generalized observables are highly discontinuous in the case of an ergodic system. 
This suggests to consider an alternative topology on the torus and with it an alternative measure. In particular, we consider a discrete topology on the configuration space (i.e.\ the torus). In such a discrete topology, every point is an open set such that the functions which are discontinuous with respect to the standard topology are actually continuous. The corresponding Hilbert space supports these functions such that we may hope to be able to represent the generalized observables on the ensuing (physical) Hilbert space -- in contrast to the situation with the previous standard quantization attempts.

\subsection{Bohr compactification of the momentum space}

Such a discrete topology can be obtained by a so--called Bohr compactification of the dual space, i.e.\ the momentum space.   Interestingly, the Bohr compactification appears as a quantization method in loop quantum cosmology \cite{Bojowald:2008zzb,Ashtekar:2011ni,Ashtekar:2003hd} or polymer quantization \cite{Hossain:2010wy,Laddha:2010hp,Corichi:2007tf} and also in the recent work on a new representation for loop quantum gravity \cite{Dittrich2014a,Bahr:2015bra} where the area operator resembles the constraint encountered here.

The Bohr compactification leads to a representation of the canonical commutation relations that is unitarily {\it inequivalent} to the usual (Schr\"odinger like) representation. The Stone-von Neumann uniqueness theorem for quantum mechanics is evaded,  as  only the exponentiated momenta can be represented on the Hilbert space, whereas the momenta themselves do not exist as operators \cite{Halvorson}.

In the following we will describe the Hilbert space resulting from such a quantization procedure. We start with the position space representations, that is functions $\psi(x_1,x_2)$ on $S^1\times S^1$.   As before, the functions are periodic $\psi(x_1+1,x_2)=\psi(x_1,x_2+1)=\psi(x_1,x_2)$.

Equipping this space with a discrete topology means also that we will have an inner product based on a discrete measure
\ba
\langle \psi_1 \,| \psi_2 \rangle &=&  \int {\rm d}\mu_{\rm d}(x_1,x_2)  \,\,\overline{\psi_1}(x_1,x_2) \psi_2(x_1,x_2)
\ea
defined such that the states
\ba\label{d2}
\psi_{\alpha_1,\alpha_2} (x_1,x_2) = \delta_{\alpha_1,x_1} \delta_{\alpha_2,x_2} 
\ea
are orthonormal
\ba\label{d3}
\langle \psi_{ \alpha_1,\alpha_2} \,| \psi_{\alpha'_1,\alpha'_2} \rangle &=&  \delta_{\alpha_1,\alpha'_1}  \delta_{\alpha_2,\alpha'_2}   .
\ea
 Note that in (\ref{d2}) and (\ref{d3}) we use Kronecker delta  functions (and not Dirac delta functions) defined as
 \ba\label{d4}
 \delta_{\alpha,\alpha'}    &=&
 \begin{cases}
 1 \q \text{for}\, \alpha=\alpha' \q \text{mod} \q 1 \\
 0\q \text{otherwise.}
 \end{cases}
 \ea
 
 The states (\ref{d2}) form an (uncountable) orthonormal basis of the Hilbert space, which is given as the closure of the set of states of the form
 \ba\label{d5}
 \psi(x_1,x_2) \,=\, \sum_{i}  c_i  \,\, \delta_{\alpha^i_1,x_1} \, \delta_{\alpha^i_2,x_2} 
 \ea
 where the sum is over finitely many indices $i$. Thus, the functions (\ref{d5}) would vanish almost everywhere (and would be highly discontinuous) with respect to the usual Lebesgue measure.
 
 Let us now pass to the momentum space representation, i.e.\ we consider  functions $\psi(k_1,k_2)$ on $\mathbb{Z} \times \mathbb{Z}$. In this momentum space representation the functions are known as {\it almost periodic functions}. The reason is that they are Fourier transforms of functions of the form (\ref{d5}). That is, we now consider (the norm--closure of the set of) functions of the form
 \ba\label{d6}
 \psi(k_1,k_2) \,=\, \sum_{i}  c_i  \,\,  \exp(- 2\pi i \, \alpha_1^i k_1) \,\,  \exp(- 2\pi i\, \alpha_2^i k_2)   .
 \ea
The inner product for these functions is defined as
\ba\label{d7}
\langle \psi_1 \,|\, \psi_2 \rangle \,=\, \lim_{T \to \infty} \frac{1}{(2T+1)^2} \sum_{|k_1|,|k_2|\leq T}  \overline{\psi_1(k_1,k_2)} \,\, \psi_2(k_1,k_2) .
\ea
and renders the (basis) states
 \ba
 \psi_{\alpha_1,\alpha_2}(k_1,k_2)\,=\,  \exp(- 2\pi i \, \alpha_1^i k_1) \,\,  \exp(- 2\pi i\, \alpha_2^i k_2) 
 \ea
 orthonormal.
 
The transformations between position space and momentum space representation are given by
 \ba
 \psi(x_1,x_2) &=& \langle \psi_{x_1,x_2}\,|\, \psi\rangle \,=\,  \lim_{T \to \infty} \frac{1}{(2T+1)^2} \sum_{|k_1|,|k_2|\leq T} \exp(2\pi i \, x_1 k_1) \,\,  \exp( 2\pi i\, x_2 k_2) \,\, \psi(k_1,k_2)\nn\\
 \psi(k_1,k_2) &=&  \int {\rm d}\mu_{\rm d}(x_1,x_2)  \,\, \exp(-2\pi i \, x_1  k_1) \,\,  \exp( -2\pi i\, x_2 k_2) \,\, \psi(x_1,x_2)   .
  \ea
 
 Note that  functions of the form $f(k_1,k_2)=\delta_{k_1,\kappa_1} \delta_{k_2,\kappa_2}$ are {\it not}  almost periodic functions and are not  elements of the Hilbert space we are considering. In fact, these functions have zero norm in the inner product (\ref{d7}).

 \subsection{Translation operators and their spectra}\label{Sec_trans}
 
This brings us to the question of how to represent the momentum operators -- and hence the Hamiltonian constraint which is a function of the momenta. Momenta are usually quantized as derivative operators, however, we now deal with discontinuous functions. Accordingly, as mentioned in the beginning of this section, the momenta themselves do not exist as operators.\footnote{This can be interpreted as a consequence of the uncertainty principle. The Hilbert space supports states that are sharply peaked in the position variables. Thus the uncertainty in the momentum would have to be unbounded, e.g the expectation values of (the square of) the momentum operator are not well defined with respect to all basis states. (The difference with the Dirac delta function state in the usual quantization scheme is that these are not elements of the Hilbert space and thus do not need to have finite expectation values with respect to observables defined on this Hilbert space.)  This situation is resolved by the fact that the momentum operator is not defined on this Hilbert space. Only the exponentiated version exist, which results in a bounded operator and hence bounded expectation values.}  Instead, their exponentiations do exist and constitute well-defined translation operators which quantize the finite (symplectic) flow generated by the momenta. 
 
 We therefore consider (in the position space representation) the translation operators
 \ba
(R_1^\mu \psi  ) (x_1,x_2) \,=\, \psi(x_1+\mu,x_2)   ,\q\q\q(R_2^\mu \psi  ) (x_1,x_2) \,=\, \psi(x_1,x_2+\mu)    .
\ea
This quantization thus introduces a novel (length) parameter $\mu$ which warrants interpretation. The momentum operator $\hat{p}_a,a=1,2$ is replaced by a $\mu$--dependent operator
 \ba
 P^\mu_a \,=\, -i \hbar \frac{1}{\mu} \left(  R_a^{\mu/2} -R_a^{-\mu/2}\right) 
 \ea
and the square of the momentum $\frac{\hat{p}^2_a}{2}$ can be quantized as
\ba
S^\mu_a :=   \frac{-\hbar^2}{ 2 \mu^2}   \left( R_a^{+\mu}  + R_a^{-\mu} - 2 \mathbb{I} \right)  .
\ea
In order to keep the formulas in this section simple, we shall work with unit masses $m_1=m_2=1$. However, there is no difficulty in working with non-unit masses and the results are qualitatively the same. For later use in section \ref{sec_screwedup}, we mention at this stage already how distinct masses $m_a,a=1,2$ can be incorporated:
\ba
S^\mu_a :=   \frac{-\hbar^2}{2\mu^2}   \left( R_a^{+\mu /\sqrt{m_a}}  + R_a^{-\mu/\sqrt{m_a}} - 2 \mathbb{I} \right) .\label{mass1}
\ea
Alternatively, one may accommodate different masses (leading in general to a different spectrum) by
\ba
(S')^\mu_a:=   \frac{-\hbar^2}{2\mu^2} \frac{1}{m_a}   \left( R_a^{+\mu}  + R_a^{-\mu} - 2 \mathbb{I} \right) .\label{mass2}
\ea

The constraint we are considering (for this section with $m_1=m_2=1$) reads
\ba
C^\mu &=& S^\mu_1+ S^\mu_2 - E.
\ea
Let us discuss the spectrum for this constraint and, more generally, for the (squared) momentum operators. This will also elucidate the meaning of the parameter $\mu$.

We start with discussing the properties -- and in particular the spectrum -- of the translation operator $R^\mu_a$. This operator is invertible, isometric and therefore unitary.  The spectrum for the translation operators has been discussed in detail in \cite{Bahr:2015bra} in the context of constructing a new representation for the loop quantum gravity observable algebra, based on a $BF$--vacuum. There, the gauge group $\rm{SU}(2)$ is also equipped with a discrete topology such that one needs to replace the angular momentum operators by (group) translation operators.

The spectrum of $R^\mu_a$ depends on whether $\mu$ is rational or irrational.  For this discussion, let us specialize to one copy of $S^1$ and a translation operator $R^\mu$; the discussion directly carries over to the torus.

~\\
\noindent \underline{$\mu = p/q\in\mathbb{Q}$ with $p,q \in \mathbb{N}$ and $\text{gcd}(q,p)=1$: } 
\\[1pt]

The Hilbert space is non--separable, however, the translation operator leaves finite dimensional subspaces invariant; e.g., a basis state $\psi_{\alpha}(x)=\delta_{\alpha,x} $ leads to the invariant subspace
\ba
\{  (R^\mu)^N \psi_{\alpha} = \psi_{\alpha- N \mu} \, |, N= 0,\ldots,q-1\}  .
\ea
This makes it straightforward to diagonalize the translation operator on each subspace. The normalized eigenfunctions are given as
\ba\label{d17}
v_{\alpha,\kappa}(x) \,=\, \frac{1}{\sqrt{q}} \sum_{l=0}^{q-1}   e^{2\pi  i \kappa l \mu }  \delta_{ \alpha+ l \mu , x } \,
\ea
with $\kappa \in \mathbb{Z}$. These have eigenvalues $e^{2\pi i\kappa \mu}$, thus $\kappa$ and $\kappa+q$ give the same value.  (Also any $\alpha'=\alpha+ l \mu$ with $l \in 0,\ldots, q$ will give the same eigenfunction, but multiplied with a phase.)  Hence, the spectrum of $R^\mu$ reads
\ba
\{ e^{2\pi i\kappa \mu} \, | \, \kappa=0,\ldots,q-1\}  .
\ea

~\\
\noindent \underline{$\mu\notin\mathbb Q$:}  \\[1pt]

We can again start with a basis vector $\psi_\alpha$ and define invariant subspaces, which now will be infinite dimensional (but separable):
\ba\label{d19}
{\cal H}^\mu_\alpha=\text{clos}(\text{span}\{ \psi_{\alpha-N \mu } \,|\, N \in \mathbb{Z}\} ) .
\ea
(Remember that the labels are to be understood as $\text{mod}\, 1$.) Note that states $\psi_\alpha$ and $\psi_{\alpha'}$ with $\alpha=\alpha'+ N \mu \, \text{mod}\, 1$ for some $N \in \mathbb{Z}$ define the same sub--Hilbert space. This defines an equivalence relation $\alpha'\sim\alpha$ on the set $[0,1)$ and we need to keep only labels in $[0,1)/\sim$.

The spectrum of $R^\mu$ on each of the ${\cal H}^\mu_\alpha$ is  continuous and given by $U(1)$. This is proven in more detail in \cite{Bahr:2015bra}. The generalized, i.e.\ non--normalizable, eigenfunctions with generalized eigenvalue $e^{2\pi i \rho} \in U(1)$  and $\rho\in [0,1)$ are given by
\ba \label{d20}
u_{\alpha,\rho}(x) \,=\, \sum_{ l \in \mathbb{Z}}  e^{2\pi i  \, l \rho}    \delta_{ \alpha + l\mu , x}
\ea
Replacing $\alpha$ with an $\alpha'\sim \alpha$ from the same equivalence class leads to a multiplication of the generalized eigenvector with a phase. 

This result can be compared with the exponentiated momenta in the usual (smooth topology) quantization: There one obtains also a spectrum $U(1)=\text{clos}\{\exp(2\pi i \mu k)|k\in \mathbb{Z}\}$ for irrational $\mu$. Yet this (seemingly continuous) spectrum is equipped with a discrete spectral measure ${\rm d}\mu_{\rm d}$ which arises from the discrete spectrum $\{k \,|\, k \in \mathbb{Z}\}$ of the momentum operator.

The compactification of the spectrum justifies the expression `Bohr compactification'.   We can  use the momentum representation (\ref{d6})  (adapted to one copy of ${\mathbb Z}$) in order to display the generalized eigenvectors.  
%
%
These are given by ($\mu$ is irrational)
\ba\label{d21}
u_{\alpha,\rho}(k)  =  
  \sum_{ l \in \mathbb{Z}}  e^{-2\pi i k \alpha} e^{2\pi i  \, l( \rho-\mu k)}   \,= \, e^{-2\pi i k \alpha} \,  \delta(\rho -\mu k)  
\ea
where $\delta(\rho-\mu k)$ denotes the Dirac delta function with respect to a continuous measure ${\rm d}\rho$ on $[0,1)$. Thus also here, the winding of the $k \in \mathbb{Z}$ values around the circle via $\exp(2\pi i k \mu)=\exp( 2\pi i \rho)$, leads to a continuous spectrum.

Knowing the spectrum of the translation operator $R^\mu$, we can construct the spectral decomposition of the Hilbert space ${\cal H}^\mu_\alpha$. (Remember that this Hilbert space is separable, since the basis (\ref{d19}) is countable.) This decomposition is based on the identity
\ba
\int_{[0,1)} {\rm d} \rho  \,\,  e^{-2\pi i l'\rho}  \,\, u_{\alpha,\rho}(x)&=&  \delta_{\alpha+ l'\mu, x} \,=:\,\psi_{\alpha+ l'\mu}(x) ,\nn
\ea
where $e^{-2\pi i l'\rho}= \langle u_{\alpha,\rho} \,|\,\psi_{\alpha+ l'\mu} \rangle$ and ${\rm d} \rho$ is the Lebesgue measure on $S^1$ and coinciding with the spectral measure. Hence, for a vector $\psi \in {\cal H}^\mu_\alpha$ we have
the spectral decomposition
\ba
\psi &=& \int_{[0,1)} {\rm d} \rho  \,\,    u_{\alpha, \rho} \,\,   \langle u_{\alpha,\rho} \,|\,\psi \rangle  \nn
\ea
into (generalized) eigenfunctions of the translation operator. This leads to a   $\rho$--representation for states in ${\cal H}^\mu_\alpha$ as follows:
\ba
\psi&=& \sum_{l \in \mathbb{Z}} c(l)   \psi_{ \alpha + l\mu } \,=\, \int_{[0,1)} {\rm d} \rho \,\,F(\rho) \, u_{\alpha,\rho} 
\ea
with
\ba
F(\rho) &=& \langle u_{\alpha,\rho} \,|\, \psi \rangle \,=\, \sum_l e^{-2\pi i l \rho} \, c(l) \q \text{and} \nn\\
c(l) &=& \langle \psi_{ \alpha + l\mu }\,|\, \psi \rangle\, =\, \int_{[0,1)} {\rm d} \rho \,\,  e^{2\pi i l \rho}  F(\rho)  \q .
\ea
The inner product between two states can now be expressed as 
\ba
\langle \psi_1 \, |\, \psi_2 \rangle &=& \sum_{l \in \mathbb{Z}} \overline{c_1(l)} c_2(l) \,\,=\,\,  \int_{[0,1)} {\rm d} \rho \,\, \overline{ F_1(\rho)} F_2(\rho) \q .
\ea
Thus, we have constructed a unitary equivalence of the super--selection sector ${\cal H}^\mu_\alpha$ to $L_2(S^1, \rm(d \rho))$, in which the exponentiated momentum operators act as multiplication operators. 
The conjugated variable, i.e.\ $x$, is represented by the discrete variable $l \in \mathbb{Z}$ via $x=(\alpha + \mu l) \, \text{mod} \, 1$.

To summarize the different representations, we had started with a non-separable Hilbert space  based on the discrete measure on the torus $S^1\times S^1$. Via a Fourier transform we could change to a Hilbert space of almost periodic functions, depending on a momentum label $k \in \mathbb{Z}$ and with a (Bohr compactification) inner product (\ref{d7}).

We then restricted to a separable Hilbert space ${\cal H}^\mu_\alpha$ on which the translation operator $R^\mu$ has a non--degenerate spectrum. This allows to use the spectral decomposition of $R^\mu$ (for irrational $\mu$) to define the $\rho$ representation, where $\rho \in [0,1)$ parametrizes the eigenvalues $\exp(2\pi i \rho)$ of the translation operator. The translation operator acts now as multiplication operator on an $L_2$ Hilbert space with compact configuration space $S^1$ and with the Lebesgue measure ${\rm d \rho}$. 

Whereas the momentum variable has been (Bohr) compactified, the configuration variable $x \in S^1$ is now parametrized by $l \in \mathbb{Z}$, where $x=(\alpha + l \mu) \, \text{mod} \, 1$. Note that due to the irrationality of $\mu$ each $l$ represents a different $x \in S^1$ value, and hence there are no periodicity requirements in the variable $l$. Again we have a winding of `lattice points' $\alpha+ \mu l$ around $S^1$, such that $S^1$ is filled densely with these lattice points.



\subsection{Spectrum of and solutions to the constraint}

The Hamiltonian (or energy operator) for a free particle on a torus in the discrete topology therefore also depends on the choice of the parameter $\mu$ and is of the form
\ba
H^\mu
\,=\, 
\left( S^\mu_1 + S^\mu_2 \right) .\nn
\ea
The operator $H^\mu$ is a linear combination of unitary (and in particular bounded) operators. Therefore, the spectrum (and `kinematical' energy) will be also bounded to be in the interval
\ba\label{d23}
\frac{\hbar^2}{ 2 \mu^2} [0, 8]   .
\ea
For $\mu=p/q$ rational the spectrum reads
\ba
\frac{\hbar^2}{ 2 \mu^2} \{   4 - 2 \cos(2\pi \kappa_1 \mu) -2\cos(2\pi \kappa_2 \mu ) | \kappa_1,\kappa_2 = 0,\ldots, q-1 \}  .\nn
\ea
For $q$ growing larger and larger the spectrum will fill out the allowed interval more and more densely. For $\mu$ irrational we indeed obtain the full interval (\ref{d23}) as spectrum. 

The bound on the spectrum gives also an interpretation to the parameter $\mu$. It can be understood as the minimal step--size. Via the bound on the Hamiltonian it also determines the maximal energy to be given as $\frac{8\hbar^2}{ 2 \mu^2}$. This is analogous to the situation in loop quantum cosmology, where the minimal step-size is related to the Planck scale and and the energy density is bounded by the Planck density \cite{Bojowald:2008zzb,Ashtekar:2011ni,Ashtekar:2003hd}.


Let us now discuss the solutions to the constraint
\ba
C^\mu=H^\mu-E ,\nn
\ea
where $E$ obviously has to be inside the interval $\frac{\hbar^2}{ 2 \mu^2} [0, 8]$. 

~\\
\noindent \underline{Solution to the constraint for $\mu\in\mathbb Q$:}  \\[1pt]

The energy of  $H^\mu$ applied to a state $v_{\alpha_1,\kappa_1} \otimes v_{\alpha_2,\kappa_2} $ is given by
\ba
\frac{\hbar^2}{ 2 \mu^2} \left(4 - 2 \cos(2\pi \kappa_1 \mu) -2\cos(2\pi \kappa_2 \mu) \right)  .\nn
\ea
Hence, for a generic $E$ there will be again no solutions to the constraints . However, with growing $q$ the allowed interval is filled more and more densely. Assuming a solution, labelled by $(\alpha_1,\alpha_2,\kappa_1,\kappa_2)$ does exist, there will be also a family of other solutions labelled by different $\alpha_1',\alpha'_2$. But these belong to different superselection sectors with respect to the translation operators $R^\mu_a$.

There is (at least) one degeneracy pattern, namely solutions with labels $(\kappa_1,\kappa_2)$ lead also to  solutions with labels $(\pm \kappa_1, \pm \kappa_2)$ (to be understood $\text{mod}\, q$)   and to $(\pm \kappa_2,\pm \kappa_1)$ for $|\kappa_1| \neq |\kappa_2|$.  It could happen that for very special energies even more solutions arise, but we expect that these cases are even more non-generic than in the smooth topology case; finding solutions in the present case requires solving a `trigonometric Diophantine problem', instead of a quadratic Diophantine equation.

As an aside, we note that allowing for states inside an energy interval $E\pm \varepsilon$ rather than of fixed $E$ -- in analogy to counting states in statistical physics -- drastically increases the solutions for sufficiently large $q$. This is related to the notion of pseudo constraints and approximate (Gaussian) projectors rather than delta-function projectors \cite{Gambini:2002wn,DiBartolo:2004cg,Gambini:2005sv,Gambini:2005vn,Dittrich:2009fb,Dittrich:2008pw,Bahr:2011xs}.   

The solutions to the constraints are normalizable, hence we can adopt the kinematical inner product as physical inner product. However, we face the same problem as with the quantization with smooth topology: Generically we have only four physical solutions. The difference is that the energy levels for which solutions do exist are now dense, but also bounded.

~\\
\noindent \underline{Solution to the constraint for $\mu\notin\mathbb Q$:}  \\[1pt]
The spectrum of $H^\mu$ is continuous such that there will be (non--normalizable) solutions as long as the energy is in the allowed interval.  $H^\mu$ has now generalized eigenvectors of the form
\ba
u_{\alpha_1,\rho_1} \otimes u_{\alpha_2,\rho_2}\nn
\ea
with generalized eigenvalue
\ba
\frac{\hbar^2}{ 2 \mu^2} \left(4 - 2 \cos(2\pi \rho_1) -2\cos(2\pi \rho_2 ) \right) \nn
\ea
and $\rho_a \in [0,1)$. The $(\alpha_1,\alpha_2)$ label equivalence classes of angles, see the discussion after eq.\ (\ref{d19}). We can restrict to one choice of $(\alpha_1,\alpha_2)$ -- this means to restrict to one superselection sector with respect to the momentum operators and thus to a separable Hilbert space. 

The solutions for a fixed (allowed) $E$ are labelled by several parameters:  We can choose the energy $0 \leq e_1 \geq E$ and as long as $e_1 \geq E- \frac{\hbar^2}{ 2 \mu^2} 4$ we can find a solution so that $H^\mu=E$. (We restrict to the generic case, i.e.\ exclude the boundary values $0$ and $\frac{8\hbar^2}{ 2 \mu^2}$ for the energy $E$.)  Additionally, there again exist  sign degeneracies in the solutions. 


The crucial point is that we now have infinitely many solutions to the constraints. 
But as the solutions are non--normalizable with respect to the kinematical inner product, we have to construct a physical inner product in which these solutions become normalizable.

\subsection{Physical Hilbert space}\label{physical HS}

We now consider the case of irrational $\mu$ and establish a physical inner product for the solutions of the constraint.  Refined algebraic quantization (RAQ) or the group averaging procedure \cite{Ashtekar:1994kv,Louko:1999tj,Marolf:1995cn,Marolf:2000iq} is unfortunately not straightforward to apply as one would have to worry about a measure for the averaging of the constraints. Similarly, for the master constraint procedure \cite{Dittrich:2004bn,Dittrich:2004bq} we would have to complete a direct integral decomposition of the Hilbert space based on the spectral measure for the constraint.

We will also -- as in the master constraint procedure -- use the spectral measure, but define the inner product similarly to the RAQ procedure. 


To this end we use the $\rho$--representation discussed in section \ref{Sec_trans}, as the constraints are diagonal in this representation. That is we also restrict to one superselection sector ${\cal H}_{\alpha_1}^\mu \otimes {\cal H}_{\alpha_2}^\mu $ with respect to the translation operators $R^\mu_1$ and $R^\mu_2$. One can thus define a formal projector onto the solutions of the constraints by
\ba
{\bf P}^\mu_{\{\alpha\}} &=& \delta\left(\tilde E  - \cos(2\pi \rho_1) -\cos(2\pi \rho_2)\right)  
\ea
where $\tilde E= 2-\frac{\mu^2}{\hbar^2}E$.  

This allows to define physical states as `projections' of kinematical states $F(\rho_1,\rho_2)$:
\ba
F_{\rm phys}(\rho_1,\rho_2) &:=&  \delta\left(\tilde E  - \cos(2\pi \rho_1) -\cos(2\pi \rho_2)\right)  F(\rho_1,\rho_2) \q .
\ea
 However as the `projector' is given by a delta--function the resulting states are not normalizable in the (kinematical) inner product of ${\cal H}_{\alpha_1}^\mu \otimes {\cal H}_{\alpha_2}^\mu $. This is due to the fact that the projector property $ {\bf P}^\mu_{\{\alpha\}} \circ  {\bf P}^\mu_{\{\alpha\}} =  {\bf P}^\mu_{\{\alpha\}}$ is not satisfied -- one rather has an infinity appearing. 

To regulate this infinity one uses nevertheless the (formal) projector property and defines (in RAQ) the physical inner product as
\ba\label{PIPD1}
\langle F_1 \,|\, F_2 \rangle_{\rm phys} &=& \langle F_1 \,|\, {\bf P}^\mu_{\{\alpha\}}\, {\cal M} \cdot F_2 \rangle \nn 
\ea
for the projections of two {\it kinematical} states.\footnote{More precisely, we would need to specify a dense subspace in the kinematical Hilbert space on which the so called rigging map, i.e.\ the projector ${\bf P}^\mu_{\{\alpha\}}$, can be well defined as an operator from this dense subspace to a (algebraic) dual space of the kinematical Hilbert space.}  Here we introduced a measure factor $  {\cal M}(\rho_1,\rho_2)$ with a positive function ${\cal M}$.


We thus have for the physical inner product 
\ba\label{physinnA}
\langle F_1 \,|\, F_2 \rangle_{\rm phys} &=&
\int {\rm d}\rho_1 {\rm d}\rho_2   \,\,  {\cal M}(\rho_1,\rho_2) \,\delta( 2-\frac{\mu^2}{\hbar^2}E - \cos(2\pi \rho_1) -\cos(2\pi \rho_2))  
\overline{F_1(\rho_1,\rho_2)} F_2(\rho_1,\rho_2) \, . \q \q
\ea

We implement the delta--function in (\ref{physinnA}) by solving for $\rho_2$
\ba
2\pi \rho_2^{\pm}(\rho_1) &=& \arccos(\tilde E-\cos(2\pi \rho_1))\nn
\ea
where we defined $\rho_2^{\pm}$ to be the roots for $\arccos$ with
\ba
\rho_2^+ \in [0,1/2) \,\, ,\q \rho_2^- \in [1/2,1)  ,\nn
\ea
(that is $\rho_2^-(\rho_1)=1-\rho_2^+(\rho_1)$) so that we have $\sin(2\pi\rho^+(\rho_1))\geq 0$ and $\sin(2\pi\rho^-(\rho_1))\leq 0$. We can then define
\ba\label{form2a}
f_+(\rho_1) := F(\rho_1, \rho^+_2(\rho_1))  ,\q\q\q f_-(\rho_2):=F(\rho_2,\rho^-_2(\rho_1))
\ea
where for $\tilde E \in [0,+2)$ we have $\rho_1\in[0, \tfrac{1}{2\pi} \arccos(\tilde E-1) ] \cup [ 1-\tfrac{1}{2\pi} \arccos(\tilde E-1),1]$. For $\tilde E \in (-2,0)$ we have $\rho_1 \in [  \tfrac{1}{2\pi} \arccos(1-\tilde E) , 1- \tfrac{1}{2\pi} \arccos(1-\tilde E)]$. (Here we take the $\arccos$--root to be in $[0,\pi)$.)

Note that the constraint equation
\ba
\cos(2\pi \rho_1) + \cos(2\pi \rho_2) &=& \tilde E
\ea
describes a deformed circle in $(\rho_1,\rho_2)$ space. $\tilde E=2-\frac{\mu^2}{\hbar^2}E$ parametrizes the circumference of this deformed circle. This circumference is small for small energy $E$ and vanishes for $E=0$. This fact can be contrasted to the quantization scheme with continuous topology: There we had a constraint of the form $h_1+h_2=E$, with $h_1$ and $h_2$ having discrete and positive spectrum. Thus one would in general expect that smaller energies lead to fewer states. This is indeed the case if one considers for instance two harmonic oscillators, where  the physical Hilbert space is always finite dimensional and the dimension increases linearly with energy (as long as these correspond to energy values leading to any  physical solution). In the discrete topology quantization of the particle on the torus this effect is now reflected in the phase space volume of the constraint surface. However the physical Hilbert space is always infinite dimensional due to the spectrum of the Hamiltonians being continuous. In the standard quantization scheme  for the particle on the torus the spectra are discrete, but the reduction of states is far more drastic than one would expect (e.g. from the harmonic oscillator example) and as discussed prevents a reasonable semi--classical limit.

In summary, physical states are characterized by a pair of functions $(f_+,f_-)$ (on the appropriate interval and with the correct periodicity behaviour in $\rho$).  Having two such states $(f_+(\rho_1),f_-(\rho_1))$ and  $(g_+(\rho_1),g_-(\rho_1))$, the physical product is given as
\ba
\langle g \,|\, f\rangle_{\rm phys} &=&
\int {\rm d} \rho \frac{{\cal M}(\rho,\rho_2(\rho)) }{2\pi |\sin(2\pi \rho_2(\rho))|}  \left( \overline g_+(\rho) f_+(\rho) + \overline{g}_-(\rho) f_-(\rho) \right)\nn
\ea
where we assume that ${\cal M}(\rho,\rho_2^+(\rho))={\cal M}(\rho,\rho_2^-(\rho))$.

As a test whether we obtain finite results for generic kinematical states, we compute the norm for the projections of the basis states $\psi_{\l_1,l_2}:=\psi_{\alpha_1 + l_1 \mu} \otimes \psi_{\alpha_1 + l_1 \mu}$. For the choice of measure factor ${\cal M}(\rho_1,\rho_2)=1$ we obtain after a variable transformation
\ba
\langle \psi_{l_1,l_2} \,| \psi_{l_1,l_2} \rangle  &=&
\frac{4}{4\pi^2}\int_{-1}^1 \int_{-1}^1 {\rm d} x_1 {\rm d} x_2 \frac{1}{\sqrt{(1-x^2_1)(1-x_2^2)}} \,\delta(\tilde E -x_1-x_2) \nn\\
&=&
\frac{4}{4\pi^2} \int_{|\tilde E|-1}^1  {\rm d} x_2 \frac{1}{\sqrt{(1-x^2_2)(1-( |\tilde E|-x_2)^2)}}   .
\ea
This gives a finite result except for $\tilde E =0$ (which one would exclude if one sticks with this physical inner product).
The result simplifies if we choose ${\cal M}= 4\pi^2 | \sin( 2\pi \rho_1)|  | \sin( 2\pi \rho_2)|$ to
\ba
4\int_{-1}^1 \int_{-1}^1 {\rm d} x_1 {\rm d} x_2 \, \delta(\tilde E -x_1-x_2) \,\,=\,\, 4(2-|\tilde E|) \q,
\ea
which is finite for all allowed values of $\tilde E$.

~\\

Remarkably, we have now obtained a physical Hilbert space in the form of an $L^2$--space over a (momentum) variable $\rho$. Note that there is the restriction for functions $(f_+,f_-)$ to be of the form  (\ref{form2a}). Here the functions $F$ from which the $(f_+,f_-)$ result are of the form (or arise as limit of functions of the form)
\ba\label{form1a}
F(\rho_1,\rho_2)  &=& \sum_{l_1,l_2} c(l_1,l_2) e^{-2\pi i l_1\rho_1}e^{-2\pi i l_2\rho_2}  . 
 \ea

 Operators on the physical Hilbert space need, in particular, preserve this form of the functions $(f_+,f_-)$. This is the case for the (exponentiated) momentum operators
\ba
e^{2\pi i \rho_1} \simeq e^{2\pi i \rho}\,,\q\q\q e^{2\pi i \rho_2} \simeq e^{2\pi i \rho^{\pm}_2(\rho)}    ,\nn
\ea
that act as multiplication operators in the $\rho$--representation.
It also holds for the following derivative operator
\ba\label{defqM}
\hat M\,:=\,
\frac{i}{2\pi}\left(
 \sin(2\pi\rho_2) \frac{\partial}{ \partial \rho_1} - \sin(2\pi\rho_1) \frac{\partial}{ \partial \rho_2} 
\right)
&\simeq& \frac{i}{2\pi}\sin(2 \pi\rho^\pm_2(\rho)) \frac{\rm d}{ {\rm d} \rho} \nn .
\ea
(Here one uses the expression with $\rho^+_2(\rho)$ for $f_+(\rho)$ and $\rho^-_2(\rho)$ for $f_-(\rho)$.)
This can be seen by evaluating the action of $\hat M$ on $F(\rho_1,\rho_2)$ explicitly:
\ba
\hat M\,\, \sum_{l_1,l_2} c(l_1,l_2) e^{-2\pi i l_1\rho_1}e^{-2\pi i l_2\rho_2}\q\q\q\q\q\q\q\q\q\q\q\q\q\q\q\q\q\q\q\q\q\q\q\q\q\q\q\q\q\q\q\q\nn\\ 
\q\q\q\q\q= 
\sum_{l_1,l_2} \frac{1}{2i} \left[ l_1(c(l_1,l_2+1)-c(l_1,l_2-1)) \,-\, l_2(c(l_1+1,l_2)-c(l_1-1,l_2))\right] e^{-2\pi i l_1\rho_1}e^{-2\pi i l_2\rho_2}\,.\nn
\ea
Note that the operator $\hat M$ annihilates the constraint (which actually is of the form $F(\rho_1,\rho_2)$ in (\ref{form1a}))
\ba
\hat M \,( \cos(2\pi \rho_1) + \cos(2\pi \rho_2) - \tilde E) &=& 0 \nn .
\ea
Therefore $\hat M$ commutes with the constraint, which acts as multiplication operator in the $\rho$--representation. Hence, one can understand $\hat M$ as providing a configuration (Dirac) observable on the physical Hilbert space. It does not commute with the exponentiated momentum $e^{2\pi i \rho_1} $ as
\ba
\left[  \hat L  \, , e^{2\pi i \rho_1} \right] \,=\, - e^{2\pi i \rho_1} \sin(2\pi \rho_2)   \nn 
\ea
which yields a non--trivial Dirac observable algebra.  This allows us to construct eigenstates and semiclassical states that minimize the uncertainty relations with respect to these observables.


%
%
%

\subsection{Physical wave functions in the kinematical configuration space}

We have constructed a physical Hilbert space and Dirac observables acting on it. Let us discuss how the physical wave functions appear in the original parametrization. That is, we want to know how the  states  $({\bf P}^{{\cal M}\mu}_{\{\alpha\}} \psi)(x_1,x_2)$ look like on configuration space.  Note that the amplitudes of these wave functions do not have a priori a well-defined meaning (in the sense of not encoding directly a probability density for the position variables, as the latter are not Dirac observables). 

\begin{center}
\begin{figure}[h]
\includegraphics[scale=0.7]{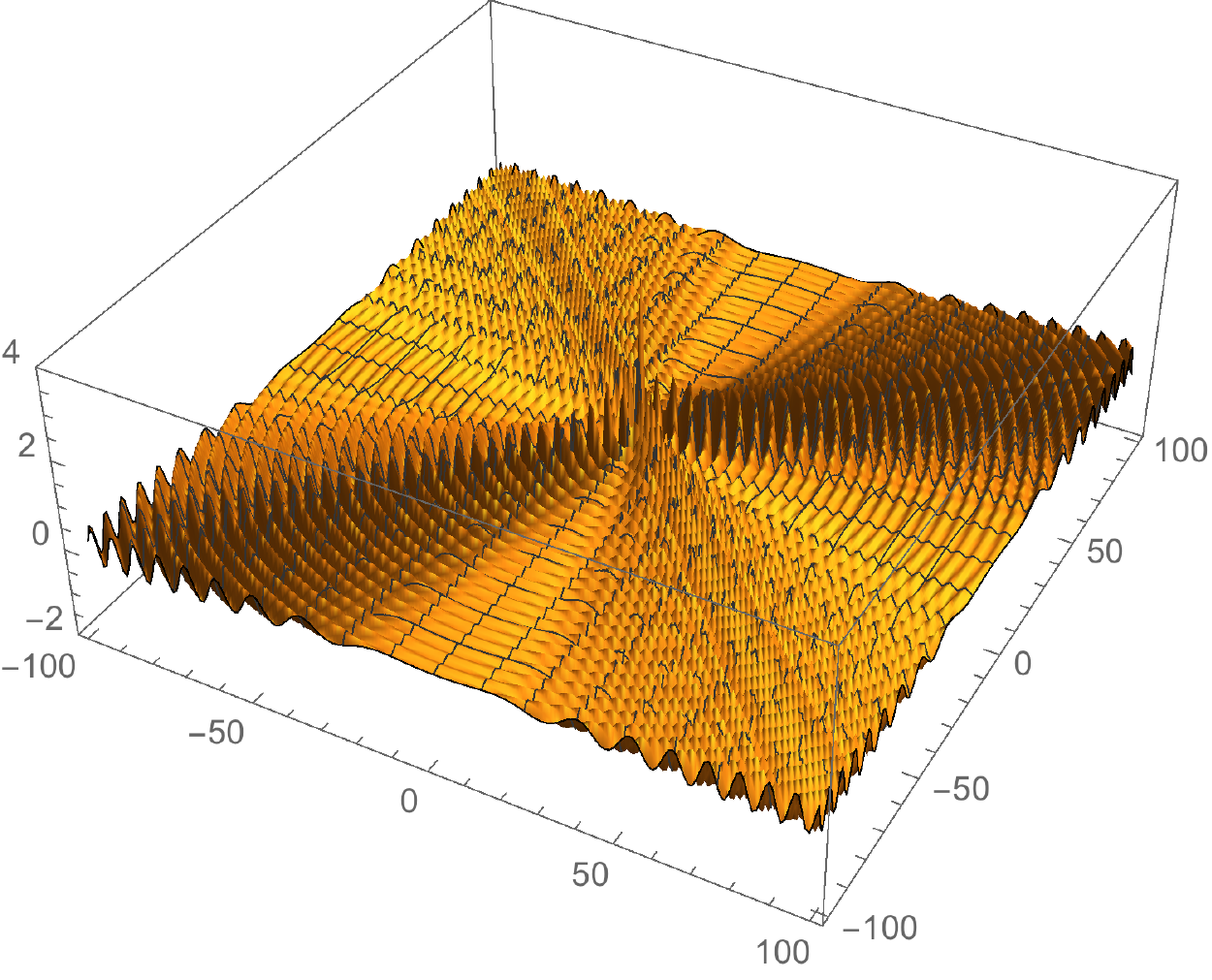}
\caption{A plot of the overlap $O(L_1,L_2)$. This also coincides with the plot of the physical wave functions obtained by projecting a basis state (peaked in the centre of the square) and not taking any winding into account.}
\label{fig:overlaps}
\end{figure}
\end{center}

The question, of course, depends very much on the initial kinematical wave function $\psi$ that is projected to a physical one.  Starting with a kinematical wave function
\ba
\psi\,=\, \sum_{l_1,l_2} c(l_1,l_2) \,\, \psi_{\alpha_1+l_1\mu, \alpha_2+l_2\mu}  \nn
\ea
(where  $\psi_{\alpha_1, \alpha_2}(x_1,x_2)=\delta_{\alpha_1,x_1}\delta_{\alpha_2,x_2}$),
the projected wave function reads
\ba
{\bf P}^{{\cal M}\mu}_{\{\alpha\}} \psi &=&
\sum_{l_1,l_2}  c_{\rm phys}(l_1,l_2) ,\, \psi_{\alpha_1+l_1\mu, \alpha_2+l_2\mu}  \nn
\ea
with
\ba
 c_{\rm phys}(l_1,l_2) &=&
 \int {\rm d}\rho_1{\rm d}\rho_2 
 \,\, {\cal M}(\rho_1,\rho_2)\delta( 2-\frac{\mu^2}{\hbar^2}E - \cos(2\pi \rho_1) -\cos(2\pi \rho_2)) \,\, \nn\\
 &&\q\q\q\q
 \sum_{l'_1,l'_2} e^{2\pi i(l_1-l'_1)\rho_1}e^{2\pi i(l_2-l'_2)\rho_2} c(l'_1,l'_2) \nn\\
& :=& \sum_{l'_1,l'_2} O(l_1-l_1',l_2-l'_2)\, c(l'_1,l'_2) \nn
\ea
where $O(L_1,L_2)$ are the overlaps between the projections of kinematical basis states
\ba\label{pinner}
O(l_1-l_1',l_2-l'_2)
&=&
\int {\rm d}\rho_1 {\rm d}\rho_2   \,\, \,\delta( 2-\frac{\mu^2}{\hbar^2}E - \cos(2\pi \rho_1) -\cos(2\pi \rho_2)) \nn\\
&&\q\q\q\q \sum_{l_1,l_2,l'_2,l'_2} \exp( 2\pi i ( (l'_1-l_1)\rho_1 + (l'_2-l_2) \rho_2))  .
\ea
  Figure \ref{fig:overlaps} provides a plot of $O(L_1,L_2)$ for $\tilde E=1$ and with the choice ${\cal M}= 4\pi^2 | \sin( 2\pi \rho_1)|  | \sin( 2\pi \rho_2)|$. Note that these overlaps $O(L_1,L_2)$ are real due to the fact that the integral splits into a sum over four (sign sector) contributions, such that the imaginary parts cancel.  $O(0,0)=4$ coincides with the physical norm of a (projected kinematical) basis state and also gives a significantly larger value than for the other $(L_1,L_2) \neq (0,0)$ values.

\begin{center}
\begin{figure}[h]
\includegraphics[scale=0.5]{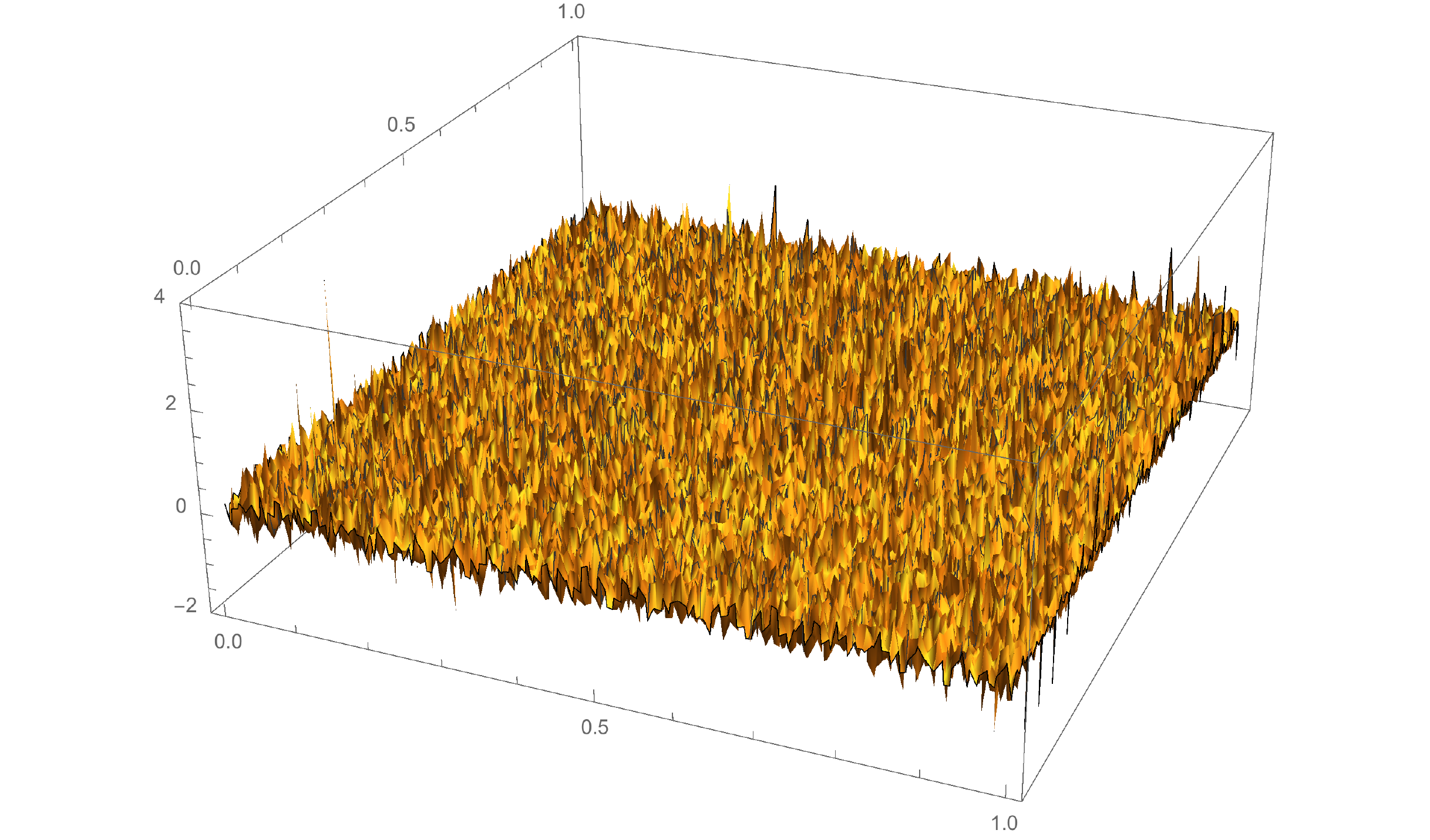}
\caption{A  physical wave function with $\mu=0.051...$ taking $200 \times 200$ values $O(l_1,l_2)$ into account, hence around 10 windings. }
\label{fig:wf1}
\end{figure}
\end{center}

Let us consider the projection of a basis function with $c(l_1,l_2)=\delta_{l_1,P_1} \delta_{l_2,P_2}$.  W.l.o.g.\ we can take $(P_1,P_2)=(0,0)$. These are (maximally) peaked (kinematical) wave functions in position space. Thus, the uncertainty in momentum is also maximal (although technically not infinite, as the momentum is exponentiated to a unitary operator). To understand the plot of such a physical wave function, we observe that $l_1,l_2$  give the position $(x_1,x_2)$ of the value $O(l_1,l_2)$ of the physical wave function where
\ba
x_1(l_1) = (\alpha_1 + l_1 \mu) \,\text{mod}\, 1\,\, ,\q x_2(l_2) = (\alpha_2 + l_2 \mu) \,\text{mod}\, 1 \nn .
\ea
Accordingly, we need to wrap the values $O(l_1,l_2)$ around the torus infinitely many times. Since $\mu$ is irrational, no point will be visited twice such that no interference effect arises. This will lead in general to a highly irregular looking wave function in the original (kinematical) position space, e.g., see figure \ref{fig:wf1}. 

This suggests that the projection of basis states will look in general irregular. A crucial point is here that the spread in momentum is maximal. The question arises if one can achieve more regular `looking' wave functions if one peaks in momentum instead. Here the difficulty is that the momentum eigenfunctions are generalized eigenvectors and thus not normalizable. As a result, we have to take into account some smearing.

The translation operator eigenfunction (just for one copy of $S^1$) in position space are given by
\ba
u_{\alpha,\rho}(\phi) \,=\, \sum_{ l \in \mathbb{Z}}  e^{2\pi i  l \rho}    \delta_{ \alpha + l\mu , \phi}  \nn.
\ea
The set $\{ (\alpha + l\mu)\, \text{mod} \,1 \, | \, l \in \mathbb{Z}\}$ will fill the interval $[0,1)$ densely. To each point in this set one associates the amplitude $a(l)=e^{2\pi i  l \rho}$.    This can give  again a highly irregular `looking' function. However, one can choose $\rho=\kappa \mu$ with $\kappa \in {\mathbb Z}$. In this case  we can consider 
\ba
e^{2\pi i \kappa \alpha} u_{\alpha,\rho}(\phi) \,=\, \sum_{ l \in \mathbb{Z}}   e^{2\pi i  \kappa( \alpha+ l\mu) }    \delta_{ \alpha + l\mu , \phi}  \nn,
\ea
which gives to nearby points in the set $\{ (\alpha + l\mu)\, \text{mod} \,1 \, | \, l \in \mathbb{Z}\}$  `nearby amplitudes'.   These cases substitute for the usual momentum eigenfunctions $e^{2\pi i \kappa \phi}$ in the standard quantization scheme.

Switching again to the torus, we will have to satisfy the constraint 
\ba
\tilde E &=& \cos ( 2\pi \rho_1) + \cos(2\pi \rho_2) \nn .
\ea
Thus, for general $\tilde E$ and a given $\rho_1=\kappa_1\mu$ we can generically not expect to find a solution $\rho_2$ of the form $\rho_2=\kappa_2\mu$. Energies which do feature solutions (in that case at least four of those) with $\kappa_1,\kappa_2 \in \mathbb{Z}$ are given by $\cos ( 2\pi \kappa_1 \mu) + \cos(2\pi \kappa_2 \mu)$. Fixing one such energy, one will have  solutions where $\rho_1$ and $\rho_2$ are of this form but also solutions where this is not the case. 

In consequence, ignoring non--normalizability, it is possible to peak on a solution of the form $\rho_1=\kappa_1\mu$ and $\rho_2=\kappa_2\mu$ which will give a `regular' looking wave function (assuming $\kappa_1,\kappa_2$ are not too large). Unfortunately, wave functions peaked exactly at a pair $(\rho_1,\rho_2)$ are neither normalizable in the kinematical nor in the physical inner product (as the latter is a $L^2$--Hilbert space with continuous measure in say $\rho_1$). One should therefore take into account a smearing function $F(\rho_1,\rho_2)$ which needs to be of the form $(\ref{form1a})$ (in particular periodic and smooth in $(\rho_1,\rho_2)$). This will in general lead to an inclusion of $\rho$--values not of the form $(\kappa_1 \mu, \kappa_2 \mu)$. 

It might still be possible to reverse engineer coefficients $c(l_1,l_2)$ such that one obtains a `regular' looking wave function.  (In particular, it would be useful to know how (physical)  eigenfunctions of $\hat L$ look like in position space.) But in any case these seem to be atypical.  This makes an interpretation of the physical wave function in terms of the transient observables \cite{Bojowald:2010xp,Bojowald:2010qw,Hohn:2011us}, mentioned in sections \ref{sec_1} and \ref{sec_standard}, difficult and thus poses a challenge to relational dynamics. Instead, it seems advantageous to interpret the dynamics in terms of physical transition amplitudes \cite{Rovelli:2004tv,Rovelli:2001bz} between kinematical states which are more generally applicable even when relational dynamics fails as in the present case. This we shall do in section \ref{semiclassD}.

\subsection{Classical interpretation}\label{classi}

Before turning to the transition amplitudes let us provide a classical interpretation of the Dirac observable found in section \ref{physical HS}. First let us comment that we use the term `(quantum) Dirac observable', as the operator $\hat M$ commutes with the quantum constraint. As we will see, however, $\hat M$ corresponds to a highly discontinuous phase space function $M$ (for which we did not use the term Dirac observable), that is indeed constant along the gauge orbits. The 'trick' that achieves the representation of $M$ in the quantum theory is the de-periodification of the $x$ variables into the $l$ parameters.  

Let us first start with the classical dynamics. Having to work with exponentiated momenta,  we replaced the  constraint
\ba
C^{0}&=& \frac{1}{2} ( p_1^2 + p_2^2)-E
\ea
with the `polymerized' expression
\ba \label{572}
C^\mu=\frac{1}{\mu^2} ( 2 - \cos(\mu p_1) -  \cos( \mu p_2))-E  \q .
\ea
$C^\mu$ approximates $C^0$ in the limit of small $\mu$ (with the approximation improving for small momenta $p_i$). However, as $p$ cannot be represented on the discrete topology Hilbert space we rather quantize $C^\mu$. There are therefore two different classical limits, one where one considers $\hbar$ to be small but $\mu$ fixed and one where one additionally considers the limit of $\mu \rightarrow 0$.

Let us consider $\mu$ fixed. In this case, the classical trajectories with respect to the constraint (\ref{572}) are given by
\ba
x_1(t) &=&  \left(\frac{1}{\mu}   \sin (\mu p_1) t + x_{10} \right) \, \text{mod}\, 1 \, , \nn\\
x_2(t) &=&  \left(\frac{1}{\mu}   \sin (\mu p_2) t + x_{10} \right) \, \text{mod}\, 1  \q .
\ea
All that changed with respect to the $C^0$ constraint is that the parameters $p_i$ are replaced by  $\tfrac{1}{\mu}\sin(\mu p_i)$. That is, the classical trajectories are the same if the momenta in the two systems are matched appropriately. Accordingly, the discussion in section \ref{sec_1} regarding the non--existence of smooth Dirac observables depending on the position variables still holds. 

Assuming for the moment that the position variables $x_i$ are in $\mathbb{R}$ instead of $S^1$ one would also replace the angular momentum Dirac observable  $L$ for the $\mu=0$ case with 
\ba
\mu L^\mu &=&   \sin(\mu p_2) x_1 - \sin(\mu p_1) x_2 \q . 
\ea
This is indeed a constant of motion for the non--compact configuration space, however with the periodic identification of the configuration space to the torus we rather have the time evolution
\ba\label{537}
\mu L^\mu(t) &=&    (\sin(\mu p_2) x_{10} - \sin(\mu p_1) x_{20} ) \,-\, (\sin(\mu p_2) n_1(t) - \sin(\mu p_1) n_2(t) ) \q .
\ea
Whereas the first term on the RHS of (\ref{537}) is constant in time, the second one is -- via the winding numbers $n_i(t)$ -- time dependent. Thus $L^\mu$ is not a global Dirac observable (but a local or transient one). 

One thus has to modify $L^\mu$ by adding the winding numbers to remove the time dependence
\ba
M' &:=&  \sin(\mu p_2)  (x_1 + n_1(x,p)) - \sin(\mu p_1)(x_2 + n_2(x,p))  \q .
\ea
Note that to define the winding numbers uniquely  we need to select for each momentum pair $(p_1,p_2)$ and each orbit labelled by this momentum pair
 a reference point $(r_1,r_2)$   on this orbit that marks the position with vanishing winding numbers.  $(n_1(x,p),n_2(x,p))$ are then defined as the number of windings along the trajectory\footnote{Note that the winding numbers would constitute an ideal clock in the sense of always going forward (here in steps) along the trajectory (although in steps of $1$, which can be remedied by adding $x_i$ to $n_i$). However to `measure' the corresponding clock one needs infinite precision in determining both positions $x_i$ and momenta $p_i$. }  starting  at $(r_1,r_2)$ and ending at $(x_1,x_2)$.  (Changing the reference points leads to a momentum dependent (and therefore invariant) term that is added to $M'$.)

 $M'$ is invariant along the trajectories as we managed to `unwind' the configuration variables $x_i$. This also applies to the parameter $l_i$ defined as
 $
 x_i=(\alpha_i + l_i \mu)\, \mod\, 1 $. If we translate the time dependence from $x_i$ to  $l_i$ (and consider these variables for the moment to be continuous) the time evolution for these variables would satisfy
 \ba\label{ctraj}
 \frac{ l_1(t_2) -l_1(t_1)}{  l_2(t_2) -l_2(t_1)} = \frac{ \sin(\mu p_1)}{\sin(\mu p_2)} \q .
 \ea
Thus the observable
\ba
M &:=& \sin(\mu p_2) l_1 \,-\, \sin(\mu p_1) l_2
\ea
is invariant under time evolution. This is indeed the classical counterpart to the quantum observable $\hat M$ in (\ref{defqM}).

\subsection{Semi-classical limit and transition amplitudes} \label{semiclassD}

To probe the semi--classical limit (at fixed $\mu$) one can consider the so--called physical  transition amplitudes \cite{Rovelli:2004tv,Rovelli:2001bz} 
\ba
W(\psi_1 \,| \psi_2) &:=&  \frac{\langle \psi_1 | \psi_2 \rangle_{\rm phys}  } {\sqrt{\langle \psi_1 | \psi_1 \rangle_{\rm phys} \langle \psi_2 | \psi_2 \rangle_{\rm phys}   }}
\ea
between kinematical states. That is, the physical  transition amplitudes are  given by the physical inner product (\ref{PIPD1}) between (the projections of) kinematical states.\footnote{The literal interpretation as transition amplitudes applies in the situation where one can chose a clock variable $T$ and if one chooses the kinematical states to include a factor $\psi_1 \sim \delta(T-\tau_1)$ and $\psi_2 \sim \delta(T-\tau_2)$.  This then gives the transition amplitude  from clock time $\tau_2$ to $\tau_1$ \cite{Dittrich:lecturenotes}. If such a clock variable $T$ is not available the physical inner product gives a generalization of the concept of transition amplitudes.} To probe semi--classical properties we can select as kinematical states semi--classical (or coherent) states, which are available more easily as compared to  constructing proper physical and semi--classical states.  A question one can consider is whether the transition amplitude between states peaked on phase space points lying on the same orbit is much larger than for states for which this is not the case.

We work in the $\rho$--representation for the (super selection sector) Hilbert space ${\cal H}_{\alpha_1}^\mu \otimes {\cal H}_{\alpha_2}^\mu $ and thus with $L_2(S^1\times S^1, {\rm d}\rho_1 {\rm d}\rho_2)$.  As semi--classical states we consider (a tensor product of) heat kernel {or complexifier} states on {$U(1)\times U(1)$} \cite{hall1994segal,hall2002coherent,winkler}, which are defined as\footnote{These are the same states as discussed in section \ref{section:Tim}. However, in \ref{section:Tim} the $U(1)\times U(1)$ gives  the (configuration space) torus, whereas here the  $U(1)\times U(1)$ is parametrized by the variables $(\rho_1,\rho_2)$ which are momentum labels.}
\ba\label{scd1}
F_{\sigma,\lambda}^t(\rho) &:=& \sum_{n \in {\mathbb Z}} e^{-\frac{t}{2}n^2} e^{n \lambda + i n 2\pi (\sigma - \rho)} \nn\\
&=& 
\sqrt{\frac{2\pi}{t}} \, e^{\frac{1}{2t}  (\lambda + i 2\pi(\sigma-\rho))^2   }
\sum_{n \in {\mathbb Z}} \exp\left(   -\frac{1}{2t} 4 \pi^2 n^2 - \frac{2\pi i n}{t} \lambda + \frac{2\pi n}{t}  2\pi ( \sigma-\rho)    \right) \q .
\ea 
Here $t$ is the classicality parameter and the states are peaked around $\rho_0=\sigma$ as well as $l_0= \lambda/t$ (as a Fourier transform to the $l$ picture shows). Furthermore, we applied in the second line of (\ref{scd1}) Poisson resummation.\footnote{ Poisson resummation uses (for suitable functions $F$) that $\sum_{n \in {\mathbb Z}}F(n \sqrt{t}) = \frac{1}{\sqrt{t}} \sum_{n \in {\mathbb Z}} \tilde F(\frac{2\pi n}{\sqrt{t}})$ where $\tilde F(k)= \int_{\mathbb R} {\rm d}y \,F(y) \exp(-iky )$ is the Fourier transform of $F(\cdot)$.} 
This changes the sum in the first line, which for small $t$ is slowly converging, to a fastly converging sum, where for sufficiently small $t$ the $n=0$ term  may give already a good approximation. (This better convergence behaviour is however traded with a loss of manifest periodicity in  $\rho$.)


Let us now consider the physical inner product
\ba
\langle F_{\sigma^o_1,\lambda^o_1} \otimes F_{\sigma^o_2,\lambda^o_2} | F_{\sigma^i_1,\lambda^i_1} \otimes F_{\sigma^i_2,\lambda^i_2}) \rangle_{\rm phys}
&=& 
\int {\rm d}\rho_1 {\rm d}\rho_2  \,\, {\cal M}(\rho_1,\rho_2) \, \delta\left(\tilde E  - \cos(2\pi \rho_1) -\cos(2\pi \rho_2)\right)   \nn\\
&&\q\q
\overline{ F_{\sigma^o_1,\lambda^o_1}^t(\rho_1) F_{\sigma^o_2,\lambda^o_2}^t(\rho_2) }  \,
F_{\sigma^i_1,\lambda^i_1}^t(\rho_1)  F_{\sigma^i_2,\lambda^i_2}^t(\rho_2) .
\ea
We can proceed by solving the delta function for e.g. $\rho_2$ and are left with an integral over $\rho_1$. If we use for the heat kernel states the Poisson resummed form in the second line of (\ref{scd1}) we can apply a saddle point approximation for the integral, valid for the limit of small $t$.  For this we consider only the $n=0$ term in each of the sums. 

It then turns out that the conditions for a saddle point {(along the real integration contour)} is that the labels $(\sigma^o,\lambda^o| \sigma^i, \lambda^i)$ describe two phase space points on the same orbit. That is, the $n=0$ term is of the form
\ba
\langle F_{\sigma^o_1,\lambda^o_1} \otimes F_{\sigma^o_2,\lambda^o_2} | F_{\sigma^i_1,\lambda^i_1} \otimes F_{\sigma^i_2,\lambda^i_2}) \rangle_{\rm phys}
&\simeq&
\sum_{\pm} \int {\rm d} \rho_1  \, \frac{4\pi^2}{t^2} g^{\pm}(\rho_1) \, \exp( \frac{1}{t} r^{\pm}(\rho_1)) 
\ea
where   
\ba\label{Rer}
\Re(r^{\pm}) &=&   \frac{1}{2}(  (\lambda^o_1)^2 + (\lambda^o_2)^2 +(\lambda^i_1)^2 +(\lambda^i_2)^2)   +\nn\\ &&
-4\pi^2 \left[ (\sigma^o_1-\rho_1  )^2 + (\sigma^o_2-\rho^\pm_2(\rho_1)  )^2+(\sigma^i_1-\rho_1  )^2+(\sigma^i_2- \rho_2^\pm(\rho_1)  )^2    \right]
\ea
and $\rho_2^{\pm}(\rho_1)=\frac{1}{2\pi} \arccos(\tilde E-\cos(2\pi \rho_1))$ denote the two roots for the $\arccos$ function. 

Assume that $(\sigma^o_1,\sigma^o_2)=(\sigma^i_1,\sigma^i_2)=:(\sigma_1,\sigma_2)$ and that $(\sigma_1,\sigma_2)$ satisfies the constraint equation (for either the $+$ or $-$ branch of the $\arccos$ root). In this case  we can easily identify the (global and local) maximum of  $\Re(r^{\pm})$ to be given by $\rho_1=\sigma_1$ and by choosing the same branch for $\rho_2$ as for $\sigma_2$. We can thus satisfy the saddle point condition for the real part of the exponent $r$. The condition on the imaginary part of $r$ is, that it has to be stationary. This imaginary part is then given as
\ba
\Im(r^{\pm}) &=&
4\pi\left[   (\lambda^i_1-\lambda^o_1)(\sigma_1-\rho_1) +(\lambda^i_2- \lambda^o_2)(\sigma_2-\rho^{\pm}_2(\rho_1))    \right]
\ea
and its derivative is
\ba
\frac{\rm d}{{\rm d}\rho_1}\Im(r^{\pm}) &=&
4\pi \left[ - (\lambda^i_1-\lambda^o_1) - (\lambda^i_2- \lambda^o_2) \frac{\rm d}{{\rm d}\rho_1} \rho_2^{\pm} \right]
\ea
with
\ba
\frac{\rm d}{{\rm d}\rho_1} \rho_2^{\pm}  =  - \frac{\sin(2\pi \rho_1)}{\sin(2\pi \rho_2^{\pm}(\rho_1))} \q .
\ea
We thus obtain the stationarity condition
\ba
\frac{ (\lambda^i_1-\lambda^o_1)}{ (\lambda^i_2- \lambda^o_2) } &=& \frac{\sin(2\pi \sigma_1)}{\sin(2\pi \sigma_2)}
\ea
which indeed reproduces the classical trajectory (\ref{ctraj}).  (Remember that the coherent states are peaked at $l=\lambda/t$.) To be precise a saddle point approximation could possibly also be applied in cases where the $\sigma$ and $\lambda$ parameters are not corresponding to a classical orbit. In this case, 
the maximum  of  the $\rho$ and $\sigma$ dependent part of $\Re(r^{\pm})$ (displayed in the second line of (\ref{Rer})) will be negative whereas for the case that the momentum label are constant and on the constraint hypersurface, the maximum vanishes. Note that the $\lambda$--dependent part of  $\Re(r^{\pm})$ is canceled by the normalization of $W$.

The expected peakedness property on the classical orbits can be confirmed by a numerical evaluation of the integrals. (We again choose ${\cal M}= 4\pi^2 | \sin( 2\pi \rho_1)|  | \sin( 2\pi \rho_2)|$.) We considered the approximation with (a) taking only the $n=0$ term into account in all sums and (b) taking the $n=0$ and $n=1$ terms into account in all sums. (This refers to the sums after Poisson resummation.) For the choice of $t=0.01$  it turns out that there is no difference  up to the order of $10^{-6}$ between the two approximations for all the transition amplitudes we computed. 

\begin{figure}
\begin{center}
\centering\includegraphics[scale=0.6]{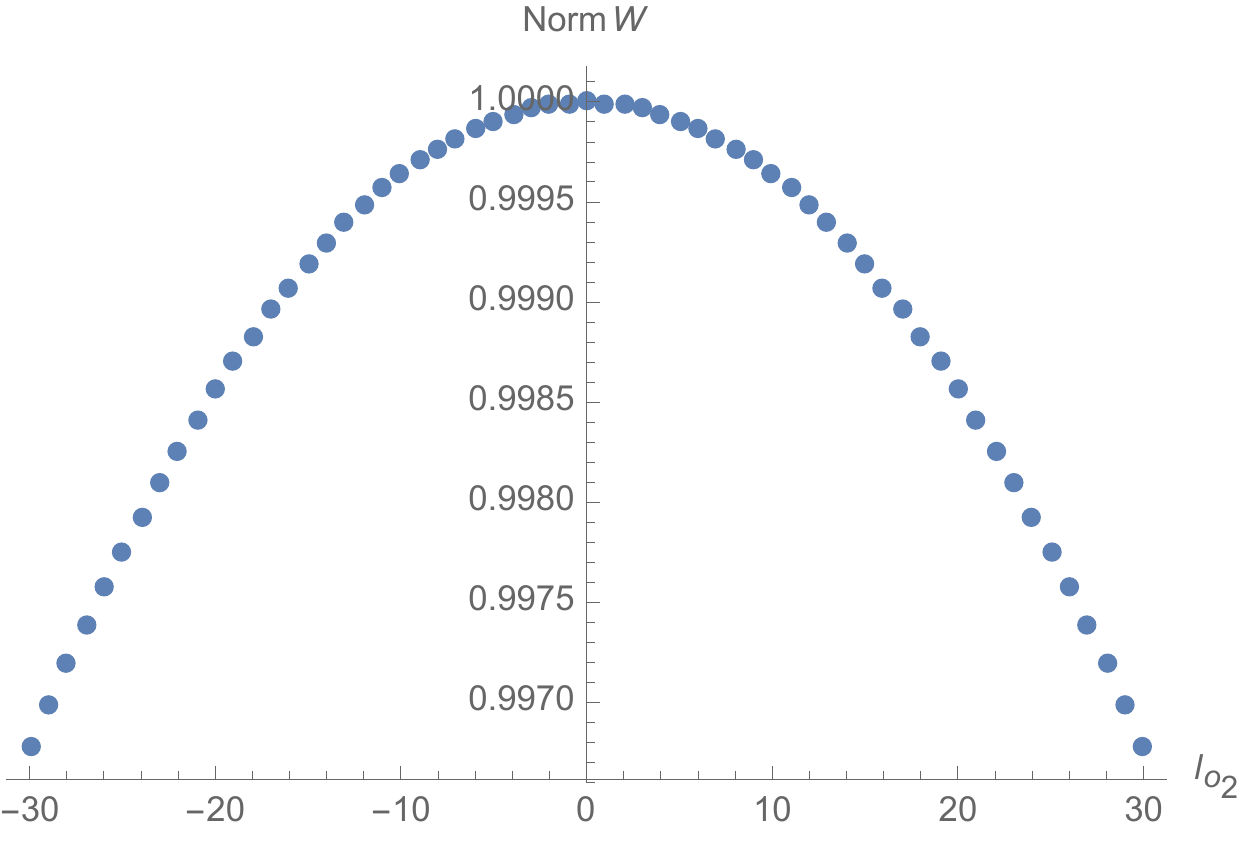}
\caption{The norm of $W$ as a function of $l_o^2$ with $l_o^1$ determined by (\ref{lcon})  and $(l_i^1,l_i^2)=(0,0)$.}
\label{fig:WL1}
\end{center}\end{figure}

Figure \ref{fig:WL1} shows the norm of the transition amplitude
\ba 
W(\sigma^o=(0.2, \rho_2^+(0.2) ), \,\, \lambda^o=  (t L_1(l^o_2), t l^o_2 )    \,\,| \,\,\sigma^i=(0.2, \rho_2^+(0.2) ),\,\, \lambda^i=(0,0))
\ea
for $\tilde E=1$ and $t=0.01$. Furthermore the function $L_2$
\ba\label{lcon}
L_1(l^o_2) := 
 \frac{  \sin(2\pi \sigma_1) }{ \sin(2\pi \rho_2^+(\sigma_1))} l^o_2
\ea
ensures that the coherent state is peaked along a classical orbit. (We can set w.l.o.g. $l^i_1=l^i_2=0$.) By varying $\lambda^o_2= t l^o_2$ we display in Figure \ref{fig:WL1} the transition amplitude `along' a classical orbit. 

\begin{center}
\begin{figure}
\center
\includegraphics[scale=0.6]{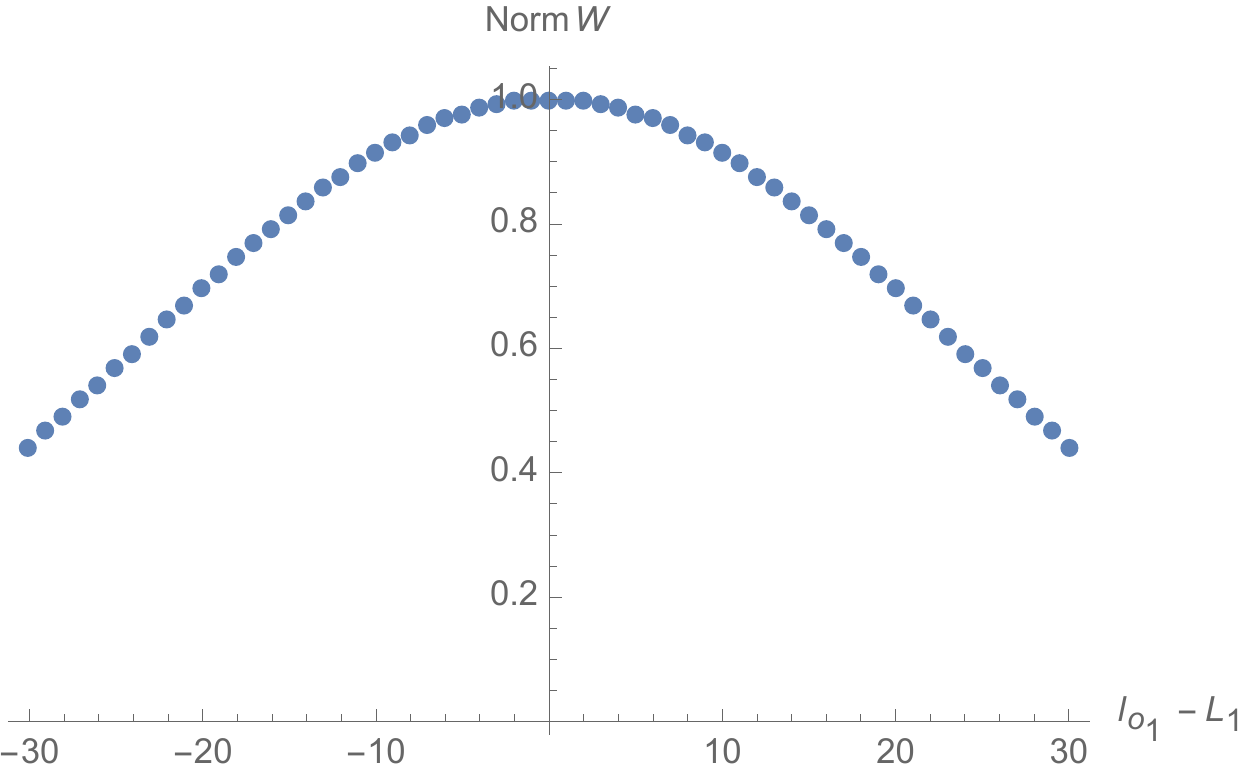}
\includegraphics[scale=0.6]{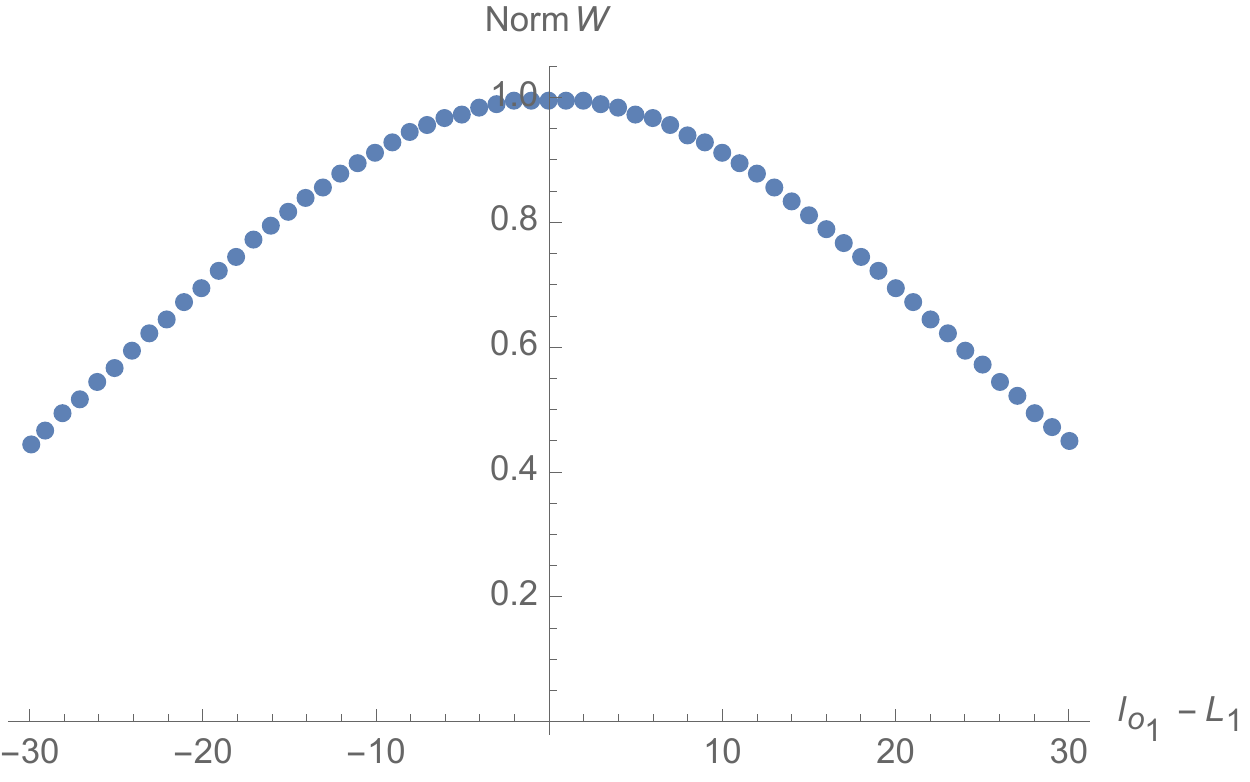}
\caption{The norm of $W$ as a function of $l_o^2$ with $l_o^1$ determined by (\ref{lcon})  and $(l_i^1,l_i^2)=(0,0)$.}
\label{fig:WL2}
\end{figure}
\end{center}

As one can see the norm of $W$  is (almost in this approximation) constant along the orbit. But the transition amplitude decays if we consider the transition amplitudes for states not satisfying the orbit condition (\ref{lcon}). Figure \ref{fig:WL2} shows the norm of W as a function of the deviation $X$ defined by  $l^o_1=L_1(l^o_2)+ X$. We choose $l_o^2=0$ in the left and $l_o^2=30$ in the right panel. In this sense we can probe the classical orbit with the physical transition amplitude involving (kinematical) coherent states. 

Note that once a choice of $\tilde E$ and $t$ is fixed (whose choice should reasonably depend on $\mu$), the parameter $\mu$ only appears implicitly. On the one hand in the translation of the $l_1$ and $l_2$ parameters into the actual positions $x_1$ and $x_2$ and on the other hand in the identification of $\rho$ with the momentum $p$ via 
$\sin(2\pi \rho)= \frac{1}{\mu}\sin(\mu p)$.   Choosing $\mu$ sufficiently small, we can  reconstruct the classical trajectories approximately for the (formally $\mu=0$) original system with constraint $C=\tfrac{1}{2}(p_1^2+p_2^2)-E$ at least for a range of the momenta.  Of course, we cannot compare to the standard quantization since as discussed  the semi--classical limit for this case is unsatisfactory. 

Again, an essential mechanism that led to these satisfactory results is the `de--periodification' of the $x_i$ variables into the $l_i$ parameters.

\section{Further toy models}\label{sec_furtherex}

One can modify the original constraint (\ref{plane}) in a number of instructive manners. All systems in this section follow from a Lagrangian analogous to (\ref{lag}) and are accordingly reparametrization invariant.

\subsection{`Cosmological' modification}\label{sec_screwedup}

To render the constraint closer to the structure of a Hamiltonian constraint in relativistic cosmology we flip the relative sign between the momenta:
\ba\label{c1a}
C=\f{p_1^2}{2m_1} - \f{p_2^2}{2m_2} - E   .
\ea
(We keep the dynamics compactified on a torus.) As before, the solutions to the equations of motion are given by (\ref{bla}), except that the second equation now features a minus sign in front of the $p_2$. We have to distinguish (1) $E=0$, and (2) $E\neq0$.

The $E=0$ case is special as it implies
\ba
\gamma\,p_1^2=p_2^2\label{equal}
\ea
and thus, as can be easily checked, if: (a) $\sqrt{\gamma}\in\mathbb Q$ all classical solutions are periodic; (b) $\sqrt{\gamma}\notin\mathbb Q$ all classical solutions are ergodic. 

~\\
 (a) The argument in section \ref{sec_1} concerning the absence of a Dirac observable which is independent of $p_1$ does not apply. Indeed, e.g.\ for $\gamma=1$, the following two functions
\ba
F_1=\sin\left(2\pi(x_1+x_2)\right)\,(p_1+p_2),\q\q\q F_2=\sin\left(2\pi(x_1-x_2)\right)\,(p_1-p_2)
\ea
are both smooth Dirac observables, independent of $p_1$ and commuting with (\ref{c1a}) everywhere on $\cc$. Notice that $F_1\equiv0$ for $p_1/p_2=-1$, while $F_2\equiv0$ for $p_1/p_2=+1$ such that both are required in order to parametrize the space of solutions, e.g.\ as $F=F_1+F_2$, which in this case will again take the form of a reduced phase space. Since everything is periodic, we can also write down relational observables as in (\ref{bla2}) (replacing $p_2$ by $-p_2$), obtaining a finite number of solutions. This model clearly is weakly integrable (see section \ref{sec_nonint} for the definition of strong, weak and reduced (non-)integrability.)

One can also quantize the model with the standard Dirac method. The solutions are given by 
\ba
\sqrt{\gamma}\,k_1=\pm k_2\,\q\q\q\q (k_1,k_2)\in\mathbb{Z}^2\label{gammasol}.
\ea
For $\sqrt{\gamma}\in\mathbb Q$ there will always be solutions to this equation. In particular, for $\gamma=1$ one obtains an infinite dimensional separable $\ch_{\rm phys}$ defined by (\ref{basis}) with
\ba
(k_1,k_2)=(0,0), (k,k)\, , (k,-k) \,  \text{with}\,\, k\in {\mathbb Z}\backslash{0} .
\ea
This Hilbert space clearly possesses sufficiently many Dirac observables to build coherent semiclassical wave packets such that the issues of the quantization in section \ref{sec_standard} do not arise.

~\\
(b) Since all orbits are ergodic, the arguments of section \ref{sec_1} apply and we thus have no continuous gauge invariant observables with non-trivial dependence on the configuration space. The system is again strongly integrable but fails to be weakly integrable.

Since there is no reduced phase space, a reduced quantization is again outright impossible. But now even a standard Dirac quantization is impossible as (\ref{gammasol}) will not have {\it any} solutions for $\sqrt{\gamma}\notin\mathbb Q$. 

However, a polymer quantization with discrete topology, as in section \ref{sec_discrete}, will again produce a non-trivial quantum theory with solutions to the quantum constraint. Replacing $p_a^2/2m_a$ by $S_a^\mu$ as in (\ref{mass1}), solutions to the constraint
\ba
C^\mu=S^\mu_1-S^\mu_2\nn
\ea
are given by
\ba
&\mu\in\mathbb Q:&\q\q\q \kappa_1,\kappa_2=0,\ldots,q-1,\q\q\q \text{s.t.}\q\q\q \cos(2\pi\kappa_1\mu/\sqrt{m_1})=\cos(2\pi\kappa_2\mu/\sqrt{m_1}),\nn\\
&\mu\notin\mathbb Q:&\q\q\q\rho_1,\rho_2\in[0,1)\q\q\q\q\q\,\,\,\q\text{s.t.}\q\q\q\,\,\,\, \cos(2\pi\rho_1/\sqrt{m_1})=\cos(2\pi\rho_2/\sqrt{m_2})
\ea
Clearly, there will always exist solutions to these equations. Alternatively, we can also replace $p_a^2/2m_a$ by $S'^\mu_a$ in (\ref{mass2}), resulting in the constraint
\ba
C'^\mu=S'^\mu_1-S'^\mu_2\nn
\ea
with a different spectrum and solutions defined by
\ba
&\mu\in\mathbb Q:&\q\q\q \kappa_1,\kappa_2=0,\ldots,q-1,\q\q\q \text{s.t.}\q\q\q \gamma\cos(2\pi\kappa_1\mu)+2(\gamma-1)=\cos(2\pi\kappa_2\mu),\nn\\
&\mu\notin\mathbb Q:&\q\q\q\rho_1,\rho_2\in[0,1)\q\q\q\q\q\,\,\,\q\text{s.t.}\q\q\q\,\,\,\, \gamma\cos(2\pi\rho_1)+2(\gamma-1)=\cos(2\pi\rho_2)
\ea
In particular, for $\mu\notin\mathbb Q$ we again obtain a continuous parameter set of solutions, resulting in an infinite dimensional non-separable $\ch_{\rm phys}$, which, however, will become separable once restricted to a superselection sector as in section \ref{sec_discrete}. Qualitatively, this Bohr compactified quantum theory will be similar to the one in section \ref{sec_discrete} such that we shall not further discuss it in detail here. The striking conclusion is: this model cannot be quantized with the reduced or standard Dirac method, but does give an interesting quantum theory once resorting to a discrete topology.

~\\
Finally, the $E\neq0$ case is qualitatively indistinguishable from the model in section \ref{sec_1}. Both resonant and non-resonant tori occur and the same arguments as before apply, showing that no Dirac observable which is independent of $p_1$ exists as a consequence of ergodicity. Similarly, no reduced phase space exists. Instead, the space of solutions is non-Hausdorff and not a manifold. However, a standard Dirac quantization will be possible, albeit resulting in a troubled quantum theory as before. These illnesses can again be remedied using polymer quantization.

\subsection{Adding a free `clock'}

Next, we can add a non-compact degree of freedom, acting as a `perfect clock' (in analogy to the highly special free massive scalar field commonly employed in loop quantum cosmology \cite{Bojowald:2008zzb,bojobuch,Ashtekar:2011ni}), such that only a subsystem will feature ergodicity. Consider the constraint
\ba
C=\f{p_t^2}{2m_t}  + \f{ p_1^2}{2m_1} + \f{p_2^2}{2m_2} - E,\q\q \q\q\q E>0 ,\label{nc}
\ea
where $t \in \mathbb{R}$ and $x_{1,2}\in[0,1)$ as before. The constraint surface is of non-compact topology $\cc\simeq \cd\times\mathbb{R}\times\mathbb{T}^2$, where $\cd$ is a disc of radius $\sqrt{E}$. The solutions to the equations of motion generated by (\ref{nc}) read
\ba
t(s)&=&\f{p_t}{m_t}\,s+t_0,\nn\\
x_1(s)&=&\f{p_1}{m_1}\,s+x_{10}-\Big\lfloor\f{p_1}{m_1}\,s+x_{10}\Big\rfloor,\nn\\
x_2(s)&=&\f{p_2}{m_2}\,s+x_{20}-\Big\lfloor\f{p_2}{m_2}\,s+x_{20}\Big\rfloor,\label{bla3}
\ea
where $s$ denotes the gauge parameter along the flow of (\ref{nc}) and $\lfloor\cdot\rfloor$ is the floor function whose values $n_1,n_2$ are the winding numbers in $x_1,x_2$ as in (\ref{bla}). Clearly, we can choose $p_1,p_2$ as Dirac observables, parametrizing $\cd$. Since these strongly commute with $C$, the system is strongly integrable. The non-trivial part of the dynamics thus takes place in the configuration space $\mathbb{R}\times\mathbb{T}^2$. We would like to be able to parametrize the trajectories by two independent Dirac observables. As in the model of section \ref{sec_1} we will have resonant tori for $\gamma\, p_1/p_2\in\mathbb{Q}$ and non-resonant tori for $\gamma\, p_1/p_2\notin\mathbb{Q}$. Consequently, in the latter case, we will again have ergodicity in $\mathbb{T}^2$. However, the important difference to the previous case is that, thanks to the first equation in (\ref{bla3}), we do not have ergodicity in $\mathbb{R}\times\mathbb{T}^2$. Consequently, the argument from section \ref{sec_1}, ruling out additional Dirac observables and a reduced phase space, does not apply to this model. Indeed, we can perform a global gauge fixing; e.g., the `clock' $T=p_tt+p_1x_1+p_2x_2$ is perfect in the sense that $\{T,C\}\approx2E>0$. Hence, $T=const$ constitutes a global gauge which is intersected by every trajectory precisely once. Accordingly, we have a global Dirac bracket and thus a reduced phase space. Every smooth function on this reduced phase space is a global, Dirac observable. This model is thus also weakly integrable.

Furthermore, it is trivial to write down relational observables with respect to either of the `clocks' $T,t$. 
For instance, choosing $t$ as the clock variable with values $\tau\in\mathbb{R}$, we obtain from (\ref{bla3})
\ba
x_1(\tau)&=&\f{m_t}{m_1}\f{p_1}{p_t}(\tau-t)+x_1-\Big\lfloor\f{m_t}{m_1}\f{p_1}{p_t}(\tau-t)+x_1\Big\rfloor,\nn\\
x_2(\tau)&=&\f{m_t}{m_2}\f{p_2}{p_t}(\tau-t)+x_2-\Big\lfloor\f{m_t}{m_2}\f{p_2}{p_t}(\tau-t)+x_2\Big\rfloor\nn.
\ea
Thanks to the floor functions, these relational observables are discontinuous and thus not Dirac observables. However, while non-differentiable in directions transversal to the constraint flow, these relational observables are differentiable along the flow, commuting with (\ref{nc}).\footnote{Note that both the winding numbers $n_1,n_2$ are just constants for a fixed value of the parameter $\tau$ as the argument of the floor function is constant.}

A reduced quantization is, in principle, possible. Also, a standard Dirac quantization can be performed producing a quantum constraint with continuous spectrum
\ba
\Big\{\f{\hbar^2 k_0^2}{2m_t}+\f{\hbar^2 k^2_1}{2m_1}+\f{\hbar^2 k_2^2}{2m_2}-  E\,\,\Big|\,\,k_0\in\mathbb R,\q(k_1,k_2)\in\mathbb{Z}^2\Big\} \nn
\ea
 thanks to $k_0 \in \mathbb{R}$. The kernel of the constraint will be parametrized by $(k_1,k_2)\in\mathbb{Z}^2$ such that $\f{\hbar^2 k^2_1}{2m_1}+\f{\hbar^2 k_2^2}{2m_2}\leq E$. Since the spectrum is continuous, one will have to construct a new physical inner product on the space of solutions. This can be done via a Klein-Gordon type inner product and taking care of the sign sectors (see, e.g., \cite{Hajicek:1993xv}). Similarly, this model can be polymer quantized. However, we shall abstain from discussing this model in further detail.
 



\subsection{Parametrizing an ergodic system}

Lastly, we briefly consider a parametrized version of the two free particles on a circle, governed by the constraint
\ba
C=p_t+\f{p_1^2}{2m_1}+\f{p_2^2}{2m_2},\label{paramc}
\ea
where $t\in\mathbb{R}$ and $x_{1,2}\in[0,1)$ as before and $\cc\simeq\mathbb{R}^3\times\mathbb{T}^2$. Ergodicity in $\mathbb{T}^2$ arises for $\gamma\,p_1/p_2\notin\mathbb{Q}$. Nevertheless, this model corresponds to an integrable unconstrained system and thus clearly features a reduced phase space and Dirac observables. More precisely, $t=\tau$ is a global gauge fixing, intersected by every trajectory. This defines a global Dirac bracket and thereby a reduced phase space. This model is thus even reducibly integrable.

The relational observables with respect to the `clock' $t$ with values $\tau$, 
\ba
x_1(\tau)&=&\f{p_1}{m_1}(\tau-t)+x_{1}-\Big\lfloor \f{p_1}{m_1}(\tau-t)+x_{1}\Big\rfloor,\nn\\
x_2(\tau)&=&\f{p_2}{m_2}(\tau-t)+x_{2}-\Big\lfloor \f{p_2}{m_2}(\tau-t)+x_{2}\Big\rfloor,\nn
\ea
are discontinuous as before, yet constant along the flow of (\ref{paramc}).

Formally, a standard Dirac quantization of this model is equivalent to reduced phase space quantization, both resulting in a Schr\"odinger equation. 
The physical Hilbert space will be $\ch_{\rm phys}\simeq L^2(S^1\times S^1)$ with sufficiently many Dirac observables. Clearly, this model can also be polymer quantized.

\section{Conclusions and discussion}\label{conc}

We have pointed out technical and conceptual challenges that arise in the context of (weakly) non-integrable constrained systems for the concept of observables and, in consequence of that, for obtaining a healthy quantum theory which admits a reasonable dynamical interpretation. Weakly non-integrable systems are devoid of differentiable Dirac observables and thereby do not give rise to a reduced phase space. Clearly, gauge invariant quantities and the space of solutions as a topological quotient space always exist, however, in contrast to the integrable case, the gauge invariant observables are non-differentiable or even discontinuous, thus not admitting Poisson-algebraic structures, and the space of solutions will not carry a symplectic structure, nor will it in general even be a manifold. This makes a quantization by the reduction method outright impossible and poses challenges to a quantization through the Dirac and path integral method since the generalized discontinuous observables cannot be represented in the quantum theory when using a standard smooth topology. At the very least, this poses challenges to the interpretation of the quantum dynamics.

This challenge is only seen if one takes the global features of the (constraint) system seriously. As mentioned, it is always possible to find sufficiently many gauge invariant functions locally around a given phase space point. It might be thus an effect which is not accessible by perturbation theory. 

Global features influence quantization significantly \cite{isham2}. Here, we have seen that the number of Dirac observables is an indicator of the size of the physical Hilbert space and thus is an important litmus test for a reasonable semi--classical limit. Dirac observables and their properties are therefore important even if one can turn to alternative tools, like the physical transition amplitudes, in order to interpret the resulting theory. Another instance where global features turn out to influence heavily the properties of the quantum theory is the spectra of (Dirac) observables \cite{spectra}. Here it is again the global flow of these observables (and thus also the global shape of the constraint hypersurface) which determines whether the spectra are discrete or continuous.

In this article, we have considered this issue only for finite dimensional systems (e.g., for cosmological models) and illustrated them with a few simple toy models. The latter suggest that there is generically a serious problem in quantizing weakly non-integrable constrained systems using standard techniques and, in particular, a smooth topology. The ensuing quantum theory -- if it exists at all -- seems to be generically ill by not possessing a semiclassical limit that connects to the classical dynamics. We stress that this problem is qualitatively distinct from the typical technical challenges arising when quantizing unconstrained non-integrable or even chaotic systems \cite{ottchaos,gutzwillerbook,berrychaos}. In the latter case, one usually at least has a short term coherence of wave packets built from superpositions over different energies (although also there wave packets will quickly run apart due to the sensitivity of initial conditions) and classically periodic orbits survive as `quantum scars' for eigenstates of sufficiently large energy. By contrast, our analysis suggests that weakly non-integrable constrained systems will generically not even have short term coherence, nor produce any classically periodic orbits (in the toy models regardless of the size of the fixed energy). In addition, also the physical transition amplitudes between kinematical states do not feature any semiclassical behaviour. These problems seem to be directly related to the scarcity of Dirac observables that can be represented in the quantum theory and, in consequence, to the resulting low-dimensionality of the physical Hilbert spaces. The key difference between the unconstrained and (totally) constrained case is that in the former one does not need to solve the dynamics in order to access the physical degrees of freedom, whereas in the latter case one does. The fact that these troubles are qualitatively distinct from the usual semiclassical subtleties of unconstrained chaotic systems is also manifested by the model of section \ref{sec_screwedup} for which a quantization by standard methods is entirely impossible, despite the system possessing well-defined classical dynamics.

We expect that the situation is qualitatively similar in the field theory case. Indeed, as discussed in section \ref{sec_nonint}, there is good evidence that full general relativity constitutes a weakly non-integrable system such that generically differentiable Dirac observables and a reduced phase space may simply not exist.\footnote{The origin of the failure of integrability of the analysed toy models was ergodicity in configuration space. A well-defined notion of ergodicity usually requires compact dynamics which one would  not expect for a typical gravitational scenario. However, non-integrability is more general than ergodicity or chaos and does not require compact dynamics.} To be sure, there exist integrable subsectors of general relativity (e.g., in the presence of special asymptotic symmetries or matter contents and in simple cosmological models), but these do not account for the generic situation. 

Classically, this does not represent a conceptual problem as one still has generalized gauge invariant observables at hand. Furthermore, as discussed in the main body, mathematically the relation between different dynamical quantities is still well-defined and gauge invariant even if the system is non-integrable. Thus, one can, in principle, still interpret the classical dynamics relationally even if the relational observables are (globally) only defined implicitly. In particular, one can resort to transient observables in the sense of \cite{Bojowald:2010xp,Bojowald:2010qw,Hohn:2011us,Hoehn:2011jw} to have a local relational interpretation along sections of a classical trajectory. As seen in the toy model of section \ref{sec_1}, one can even explicitly solve for such transient relational observables on branches of the solution (see also \cite{Hohn:2011us} for a discussion of this in a chaotic cosmological model). Thus, locally predictions and explicit relations are possible and this better be the case as this is the situation in our own observable (spatial and temporal) vicinity of the universe. For example, also our solar system is chaotic \cite{arnold2007mathematical}, but nevertheless for time scales relevant to us we can actually compute the orbits of celestial bodies to a high accuracy.

However, in the quantum theory it is difficult to make sense of relational dynamics in the context of non-integrability -- at least in the usual sense of locally scanning through the a priori timeless physical quantum state. For systems featuring the `global problem of time', a given value of a relational `clock' reading admits multiple classical solutions for the evolving degrees of freedom. States in the deep quantum regime will overlap on the classically distinct solutions and superpose internal time directions such that it is unclear how to interpret this relationally \cite{Bojowald:2010xp,Bojowald:2010qw,Hohn:2011us}. This becomes especially challenging for systems where a given clock value admits densely many classical solutions. For example, in the discussed toy models the physical quantum states can be uniformly spread out (or, in the discrete topology, peaked almost everywhere) on the torus. It is not clear how to interpret this in terms of the relational paradigm. This is qualitatively similar to the analysis in \cite{Hohn:2011us}, showing that the effective relational dynamics in the closed FRW model with minimally coupled massive scalar field breaks down generically in the region of chaotic scattering.

The example of a particle on a torus illustrates the interpretational challenge well: suppose that this model described a cosmological dynamics and that the time that the particle needs to traverse the torus once is a cosmologically large time, e.g.\ corresponds to one period in an (almost) cyclic universe. The problem of not having sufficiently many Dirac observables would only arise if one averages over the cycles, however not if one applies a more `local' quantization approach.

Do weak non-integrability and chaos thus constitute a fundamental obstruction for canonical quantum gravity? Not necessarily. While they pose challenges to quantizing general relativity and interpreting the resulting quantum theory, the present manuscript does offer a possible resolution. One may exploit the fact that the act of quantization is a rather ambiguous procedure. This, in particular, applies if the (physical) phase space features a non--standard topology, or as in the examples considered here is not  a manifold or not even Hausdorff.  It is important for gravity, as the physical phase space for general relativity will certainly not be given by a manifold. The key idea for the examples considered here is to modify the quantization procedure by refining the topology of the configuration space until the discontinuous generalized observables become continuous in the new topology and can be represented in the quantum theory.  A similar strategy might work in the full theory (assuming an overwhelming amount of technical challenges can be overcome): one has to choose the quantization of the kinematical Hilbert space such that (a) the imposition of constraints leads to sufficiently many states allowing for a reasonable semi--classical limit and (b) (which as we showed is related to (a)), the resulting physical Hilbert space allows for a representation of (sufficiently many) gauge invariant variables.

Here we have employed a discrete topology which leads to a Bohr compactification of momenta in the quantum theory. This corresponds to polymer quantization \cite{Hossain:2010wy,Laddha:2010hp,Corichi:2007tf} which is the quantization technique of loop quantum cosmology \cite{Bojowald:2008zzb,Ashtekar:2011ni,Ashtekar:2003hd} and of a new representation of loop quantum gravity \cite{Bahr:2015bra}. Remarkably, polymer quantization has remedied the illnesses of the standard quantization techniques and produced a non-trivial quantum dynamics with sufficiently many observables in the toy models of this article. The price to pay for this is that the quantum theory becomes more challenging as one now has to deal with highly discontinuous observables and states. But this is a technical problem and not a problem in principle. Furthermore, one has to deal with super-selection sectors and to interpret the additional quantization parameter which, however, in a gravitational context can be related to the Planck scale \cite{Bojowald:2008zzb,Ashtekar:2011ni,Ashtekar:2003hd}.  Note that this additional parameter arises from the necessity to use exponentiated momenta, which in turn can be understood as a consequence of the uncertainty relations and the choice to have states sharply peaked in the configuration variable as normalizable states. The latter choice is motivated by the desire to represent discontinuous observables in the quantum theory.

 While the present work provides non-trivial evidence that loop quantum gravity inspired quantization techniques give rise to an interesting quantum theory, it is of course an open question whether these techniques will also work for full general relativity -- and ultimately whether our own universe is actually described by such a modified quantum theory.

We emphasize that the challenges to a fundamentally relational interpretation of the quantum dynamics mentioned above remain the same for the polymer quantization. But this is not a dramatic observation because one may resort to an interpretation in terms of physical transition amplitudes between kinematical states as proposed in \cite{Rovelli:2004tv,Rovelli:2001bz}. Since the physical transition amplitudes are equivalent to the physical inner product, they are always applicable (whenever the physical Hilbert space exists) and indeed retain a meaning in the presence of chaos when the usual relational interpretation seems to fail.

Before closing, we mention two alternatives to polymer quantization. Firstly, one may modify the constraint quantization technique by, instead of requiring the physical Hilbert space to be solely defined through the constraint kernel, also allowing states which are `almost' annihilated by the constraints. One might attempt to motivate this operationally from the fact that it is fundamentally impossible to measure physical quantities exactly. This modification leads to the notion of pseudo-constraints and approximate rather than exact delta function projectors \cite{Gambini:2002wn,DiBartolo:2004cg,Gambini:2005sv,Gambini:2005vn,Dittrich:2009fb,Dittrich:2008pw,Bahr:2011xs}. For instance, in the toy models the constraint has a form $C=H-E$. Rather than requiring that the energy be exactly $E$, one could permit values in an interval $E\pm\epsilon$ in analogy to counting states in statistical physics in small energy intervals rather than at exact energy values. Technically, this could be achieved by changing the constraint to $\tilde{C}=\epsilon\lfloor H/\epsilon\rfloor - E$ which generically leads to a larger kernel (and thus richer quantum theory) than the original $C$. 
While statistical properties of the constraint spectrum are unimportant when working with exact constraints (except whether the spectrum is discrete or continuous around zero), they will become useful for pseudo constraints. For example, in this case periodic orbit theory and concepts such as the Gutzwiller trace formula \cite{gutzwillerbook,gutzwiller1971periodic} may be applicable to determine the density of states in the `vicinity' of the original constraint. 

Related to this, it is interesting to study a path integral quantization of the examples considered here. Path integral quantization for constraint systems is deeply related to Dirac quantization, as the path integral acts as a projector onto the solutions space of the constraints \cite{Halliwell:1990qr,Rovelli:1998dx}. However, this and the resulting solution space depends strongly on the choice of the measure in the path integral. For example, a standard procedure is to use a discretization of the path integral, which leads typically to a breaking of the (diffeomorphism) symmetry which is deeply intertwined with the (Hamiltonian) constraints \cite{BahrDittrich09}. The resulting path integral just imposes the pseudo-constraints discussed above, see also the discussion in \cite{uniform-vs-continuumlimit}.  More generally, one usually interprets the path integral as a procedure which requires less global information, and it would thus be interesting to see whether and how one can reproduce the results of the Dirac quantization discussed here.  As the problem of having only discontinuous observables arises from averaging the (ergodic) dynamics over (infinitely large) time intervals, the issue discussed here might indeed be connected to `infrared issues' that result from discussing minisuperspace path integrals and related probability interpretations for infinitely large universes (see also the recent works \cite{Shenker, Marolf15}).

Lastly, it is worth mentioning that there exist modified gravity theories which do {\it not} face the problem of non-integrability as discussed in this article. For example, neither shape dynamics \cite{Gomes:2010fh,Barbour:2013goa,Mercati:2014ama}, nor Horava-Lifshitz gravity \cite{Horava:2009uw,Donnelly:2011df}, nor Einstein-aether theory \cite{Jacobson:2008aj} possess a Hamiltonian constraint.\footnote{See also \cite{UnruhWald} which suggest a version of uni-modular gravity to address this problem (at least for quantum cosmology).} Instead, they feature a proper Hamiltonian. Accordingly, in these theories one does not need to solve the dynamics in order to access the physical degrees of freedom such that chaos does not pose a fundamental challenge.


%

\section*{Acknowledgements}
BD thanks Benjamin Bahr for discussions on the spectrum of the translation operator and PH would like to thank Sylvain Carrozza and Aldo Riello for discussion. We furthermore thank Ted Jacobson, Carlo Rovelli and Bill Unruh for enlightening remarks. Research at Perimeter Institute is supported by the Government of Canada through Industry
Canada and by the Province of Ontario through the Ministry of Research and Innovation. MN is grateful
to AIMS Ghana for a stipend and was also supported by an NSERC grant awarded to BD. TK received financial support from the
Foundational Questions Institute (fqxi.org) and thanks the Perimeter Institute for its hospitality during the start of this work.


\bibliography{bibliography}{}
\bibliographystyle{utphys}

\end{document}